\documentclass[11pt]{article}
\usepackage{amsmath,epsfig,amssymb,fancyhdr,graphicx,wrapfig,multirow,multicol,lscape}
\usepackage[top=1in, bottom=1in, left=1in, right=1in]{geometry}
\pdfoutput=1

\newcommand{\var}[1]{\mathsf{Var}({#1})}
\newcommand{\mean}[1]{\langle{#1}\rangle}
\newcommand{\abs}[1]{\left|{#1}\right|}
\newcommand{\prob}[1]{\mathbb{P}({#1})}

\begin{document}


\title{Needles in the Haystack:\\ Identifying Individuals Present in Pooled Genomic Data}


\author{Rosemary Braun, William Rowe, Carl Schaefer\\ Jinghui Zhang, and Kenneth Buetow\\ {\footnotesize{\textit{National Cancer Institute, NIH, Bethesda, MD.}}}}

\date{\today}

\maketitle

\begin{abstract}

Recent publications have described and applied a novel metric that
quantifies the genetic distance of an individual with respect to two
population samples, and have suggested that the metric makes it possible
to infer the presence of an individual of known genotype in a sample
for which only the marginal allele frequencies are known. However,
the assumptions, limitations, and utility of this metric remained
incompletely characterized.
Here we present an exploration of the strengths and limitations of
that method. In addition to analytical investigations of the underlying
assumptions, we use both real and simulated genotypes to test empirically
the method's accuracy.  The results reveal that, when used as a means
by which to identify individuals as members of a population sample,
the specificity is low in several circumstances. We find that the
misclassifications stem from violations of assumptions that are crucial
to the technique yet hard to control in practice, and we explore the
feasibility of several methods to improve the sensitivity.  Additionally,
we find that the specificity may still be lower than expected even
in ideal circumstances.  However, despite the metric's inadequacies
for identifying the presence of an individual in a sample, our results
suggest potential avenues for future research on tuning this method to
problems of ancestry inference or disease prediction.
By revealing both the strengths and limitations of the proposed method, we
hope to elucidate situations in which this distance metric may be used in
an appropriate manner. We also discuss the implications of our findings in
forensics applications and in the protection of GWAS participant privacy.

\end{abstract}


\section{Introduction}
In the recently published 
article ``Resolving Individuals Contributing Trace Amounts
of DNA to Highly Complex Mixtures Using High-Density SNP Genotyping
Microarrays''~\cite{HOME08}, the authors describe a
method by which the presence of a individual with a known genotype may
be inferred as being part of a mixture of genetic material for which
marginal minor allele frequencies (MAFs), but not sample genotypes, are known.

The method~\cite{HOME08} is motivated by the idea that the presence of a specific
individual's genetic material will bias the MAFs of a sample of
which they are part in a subtle but systematic manner, such that when
considering multiple loci, the bias introduced by a specific individual
can be detected even when his DNA comprises only a small fraction of
the mixture.  More generally, it is well known that samples of a
population will exhibit slightly different MAFs due to sampling variance
following a binomial distribution; the genotype of the individual in
question contributes to this variation, and so may be ``closer'' to a
sample containing him than to a sample which does not.  Based on
this intuition, the article~\cite{HOME08} defines a genetic distance statistic
to measure the distance of an individual relative to two samples,
summarized as follows:

Consider an underlying population $P$ from which two samples 
$F$ (of size $n_F$) and $G$ (of size $n_G$) are drawn independently
and identically distributed (i.i.d.) [in~\cite{HOME08}, these are referred to as ``reference'' and ``mixture'' respectively].  Consider now an additional sample $Y$; we wish to detect
whether $Y$ was drawn from $G$, versus the null hypothesis that 
$Y$ was drawn from $P$ independent of $G$ and $F$.   
Given the MAFs $f_i$ and $g_i$ at locus $i$ for $F$ and $G$, respectively, and given the MAFs $y_i$ for sample $Y$ with  $y_i \in \{0,0.5,1\}$
(corresponding to homozygous major, heterozygous, and homozygous minor
alleles)
at each locus $i$, \cite{HOME08} defines the relative distance of sample $Y$
from  $F$ and $G$ at $i$ as:
\begin{equation}
D_i (Y) =  \abs{y_i - f_i} - \abs{y_i - g_i} \, .
\label{craigD}
\end{equation}
By assuming only independent loci are chosen and invoking the central
limit theorem for the large number of loci genotyped in modern studies,
the article~\cite{HOME08} asserts that the $Z$-score of $D_i$ across all loci 
will be normally distributed,
\begin{align}
T(Y) &= \frac{\mean{D_i} - \mu_0}{\sqrt{\var{D_i}/s}} 
     = \frac{\mean{D_i}}{\sqrt{\var{D_i}/s}} \sim N(0,1)
\label{craigT}
\end{align}
where $\mean{\cdot}$ denotes the average over all SNPs $i$,
$s$ is the number of SNPs, and Eq.~\ref{craigT} exploits the 
assumption~\cite{HOME08} that an individual who is in neither $F$ nor $G$ will
be on average equidistant to both under the null hypothesis, i.e., $\mu_0 =0$. 
The article~\cite{HOME08} proposes using this approach in a forensics
context, in which $G$ is a mixture of genetic material of unknown
composition (e.g., from a crime scene), and $Y$ is suspect's
genotype; by choosing an appropriate reference sample for group
$F$, it is hypothesized that large, positive $T$ will be obtained for
individuals whose genotypes are included in $G$, and hence bias $g_i$,
while individuals whose genotypes are not in $G$ should have insignificant
$T$ since they should intuitively be no more similar to the mixture sample $G$ than 
they are to the reference sample $F$.

In~\cite{HOME08}, the authors applied this test to a multitude of individuals
$Y$, 
each of which are present in the samples constructed by them for $F$ or $G$,
and report near-zero false negative rates.  The article concludes that 
it is possible to identify the presence of 
DNA of specific individuals within a series of highly complex genomic
mixtures, and that these ``findings show a clear path for
identifying whether specific individuals are within a study based on
summary-level statistics.'' In response, many GWAS data sources have
retracted the publicly available frequency data pending further study of
this method due to the concern that the privacy of study participants
can be compromised.  However, because no samples absent from both $F$
and $G$ were used, false positive rates---significant $T$ for individuals
neither in $G$ nor $F$---are not assessed in practice; rather, they
are simply assumed to follow the nominal false-positive rate $\alpha$
given by quantiles of the putative null distribution in Eq.~\ref{craigT}.

In this manuscript, we expand on~\cite{HOME08} by investigating
the method's robustness to several inherent assumptions:
\begin{enumerate}
\item that $F$, $G$, and $Y$ are all i.i.d. samples of the same population $P$ and hence the difference of MAFs $f_i$ and $g_i$ in the two samples is small;
\item that the loci $i$ are independent, such that the central limit theorem may be invoked in Eq~\ref{craigT}; and
\item that an individual $Y_-$ in neither $G$ nor $F$ does not have sufficient genotype identity (e.g., via inheritance) to true positive individual $Y_+$ that $D_i (Y_-) \approx D_i(Y_+)$ for enough $i$ to bias $T(Y_-)$. 
\end{enumerate}
To investigate the effect of these assumptions, 
we begin with a statement of the problem
that~\cite{HOME08} attempts to address, analytically derive the
effect of deviations from the assumptions, and empirically explore
the accuracy of the method in practice using real and simulated
genotype data.  We conclude with a discussion of the implications
of our findings, both in forensics as well as regarding 
identification of individuals contributing DNA in GWAS.

The results presented here reveal that membership classification via
Eq.~\ref{craigT} is sensitive to the choice of  $F$ and $G$; that even a
small amount of LD will alter the distribution of $T$ for null samples;
and that individuals who are related to
members of $F$ or $G$ are frequently assigned significant $T$ values.
Our findings suggest that Eq.~\ref{craigT} will in practice yield a high
false-positive rate if used to discern the membership of an individual
in a specific sample, and when used for this purpose is likely be accurate
only if the above assumptions are exceedingly well-met and the individual $Y$
is believed \textit{a priori} to be present in exactly one of  $F$
or $G$.  However, although these findings suggest that Eq.~\ref{craigT}
may have limited utility to reliably detect the identity of an
individual in $F$ or $G$ without prior knowledge, it may be valuable
for verifying that an individual is \textit{not} in
either sample, and we find some suggestion that the metric 
(Eq.~\ref{craigD}) proposed in~\cite{HOME08} could perhaps be
extended to other genetic-similarity problems  
(e.g., in ancestry inference).


\section{Materials and Methods}

We explore the performance of the method described in~\cite{HOME08} both
analytically and empirically. For the empirical studies, we attempt
to classify real
and simulated samples into pools derived from publicly available data
sources in order to assess the chances that an individual is mistakenly
classified into a group which does not contain his specific genotype.  The
data used in these tests is described below:

\subsection{Experimental genotypes and MAFs\label{truth1}}
Real-world genotypes from publicly available data sets were retrieved
as follows: 2287 samples with known genotypes were obtained from the
Cancer Genomic Markers of Susceptibility (CGEMS) breast cancer study.
The samples were sourced as described in~\cite{CGEMS07}.  Briefly,
the samples comprised 1145 breast cancer cases (sample group C+) and
a comparable number (1142) of matched controls (group C--) from the
participants of the Nurses Health Study.  All the participants were American
women of
European descent.  The samples were genotyped against the Illumnina
550K arrays, which assays over 550,000 SNPs across the
genome.  To assess the genetic identity shared between  samples, we computed the
fraction of SNPs with identical alleles for all possible pairs of individuals;
none exceeded $0.62$.

Additionally, 90 genotypes of individuals of  European descent
(CEPH) and 90 genotypes of individuals of Yoruban descent (YRI) were 
obtained from the HapMap Project~\cite{HAPMAP03}.  In both cases,
the 90 individuals were members of 30 family trios comprising
two unrelated parents and their offspring.  SNPs in common
with those assayed by the CGEMS study and located on chromosomes 1--22
were kept in the analysis (sex chromosomes were excluded since the
CGEMS participants were uniformly female); a total of 481,482 SNPs met
these criteria.

Minor allele frequencies for case and control groups were
computed from the CGEMS genotypes.  Publicly-available minor
allele frequencies from the 60 unrelated CEPH individuals 
were retrieved directly from the HapMap Project~\cite{HAPMAP03}.  The 
distribution of MAF differences for each group may be seen in 
Fig.~\ref{freqdists}.

\subsection{Simulated Genotypes I\label{sim1}} 
To explore the potential for a sample whose genotype is drawn on $f_i$ or $g_i$  (without being a member of $F$ or $G$) to
be misclassified, 
five sets of 320 simulated genotypes were created by drawing a genotype for
each SNP independently 
as a pair of Bernoulli trials from given allele frequencies:
\begin{itemize}
\item[\textbf{S.1:}] For each locus in each sample, genotypes were drawn on the CGEMS control allele frequencies for that locus.
\item[\textbf{S.2:}] For each locus in each sample, genotypes were drawn on the CGEMS  case allele frequencies for that locus.
\item[\textbf{S.3:}] For each locus in each sample, genotypes were drawn on the HapMap CEPH~\cite{HAPMAP03} allele frequencies for that locus.
\item[\textbf{S.4:}] For each sample, 50\% of the loci were selected at random to have genotypes drawn on CGEMS case frequencies, and the other 50\% had genotypes drawn on CGEMS control frequencies.
\item[\textbf{S.5:}] For each sample, 50\% of the loci were selected at random to have genotypes drawn on HapMap CEPH frequencies, 25\% of the the of the loci were selected at random to have genotypes drawn on CGEMS case frequencies, and the other 25\% had genotypes drawn on CGEMS control frequencies.
\end{itemize}

\subsection{Simulated Genotypes II\label{sim2}}
To further explore the influence of genetic similarity, two other
simulation sets were created.  Beginning with the MAFs from CGEMS controls, 
here denoted by $p_i$, we create the first set as follows:
\begin{enumerate} 
\item Draw $f_i$ from $\mathsf{Bin}(2000, p_i)/2000$ to simulate the MAFs of a sample of 1000 individuals;
\item Draw 1000 genotypes on $\mathsf{Bin}(2, p_i)/2$ to simulate genotypes of 1000 individuals who will comprise $G$;
\item Construct 200 genotypes ($Y$s) for which $q$ percent of SNPs are chosen at random to be identical to a specific $G$ individual (selected at random for each of the 200 samples), and the other $1-q$ fraction  of SNPs are drawn on $\mathsf{Bin}(2, p_i)/2$;
\item Perform step 3 for values of $q$ in 0.01 increments from 0 to 1, thus generating 100 pools of 200 samples each who bear $q$ identity to a true-positive individual, and apply Eqs.~\ref{craigD},\ref{craigT} to classify them against the $F$ and $G$ generated in steps 1 and 2.
\end{enumerate}
A second set is created as follows, also using the MAFs from CGEMS controls as $p_i$:
\begin{enumerate}
\item Draw $f_i$, $g_i$ independently from $\mathsf{Bin}(2000,p_i)/2000$ to simulate the MAFs of two samples of 1000 individuals each;
\item Draw 200 genotypes ($Y$s) on $\mathsf{Bin}(2, (1-q) p_i + (q)g_i)/2$ to simulate 200 individuals from a population with MAFs biased toward $G$ by $q$ percent;
\item Perform step 2 for values of $q$ in 0.01 increments from 0 to 1, thus generating 100 pools of 200 samples each to be classified against the $F$ and $G$ generated in step 1. 
\end{enumerate}
By creating these sets, we ensure that we have samples for which all SNPs are independent in $F$ and $G$, and that $F$ and $G$ are samples of the same
underlying population; the classification can then be observed as a function of
the similarity parameter $q$ in both cases.

\subsection{Classification of real and simulated genotypes}
The method as described in~\cite{HOME08} and summarized in the
Introduction was implemented using R~\cite{R}.  Subsets of the real data
(Sect.~\ref{truth1}) and simulated data (Sect.~\ref{sim1}) described
above were classified in a total of 17 tests, starting with a total of 481,382 SNPs and excluding those which did not achieve a minor allele frequency $>$0.05 in both
$F$ and $G$ for a given test.  A summary of the tests is
provided in Table~\ref{testtab}.  Additionally, a series of 200 tests
using $Y$, $f_i$, and $g_i$ as described in Sect.~\ref{sim2} were performed.
 

\section{Results\label{allres}}
We begin with an analytical exploration of the assumptions underlying Eq.~\ref{craigD},\ref{craigT}, followed by the results of the tests as described in Methods. 

\subsection{$D_i$ and $T$ under the null hypothesis \label{anal}}

To address the need for a fully rigorous examination of the problem
which~\cite{HOME08} tries to address, we here attempt to set up an
idealized situation to which the theory and methods in~\cite{HOME08}
apply, and consider the properties of $D_i$ and $T$
(Eqs.~\ref{craigD},~\ref{craigT}) in that setting versus 
deviations from that setting.  

Let us assume an underlying population $P$ with MAFs $p_i$ from
which samples $F$ (of size $n_F$) and $G$ (of size $n_G$) are drawn
i.i.d.  Consider now an additional sample $Y$.  The null hypothesis
is that $Y$ was drawn from $P$, independent of $F$ and $G$; the
alternative of interest is that $Y$ is drawn from $G$ (or,
symmetrically, $F$).
Under these idealized circumstances, we observe that:
\begin{eqnarray}
f_i  &\sim& \mathsf{Bin}(2n_F, p_i)/2n_F \, , \label{idealf}\\
g_i  &\sim& \mathsf{Bin}(2n_G, p_i)/2n_G \, , \label{idealg}\\
y_i &\sim& \mathsf{Bin}(2, p_i)/2 \, ,
\label{idealy}
\end{eqnarray}
where the factors of two are a consequence of each sample possessing
two independent alleles per locus.  In~\cite{HOME08}, it is proposed that $T$
(the $Z$-score of $D_i$ across all SNPs) follows a standard normal
distribution (Eqs.~\ref{craigD},\ref{craigT}).  This proposition rests 
upon two assumptions: namely,
that the mean $\mean{D_i}$ across all SNPs under the null hypothesis
is zero, i.e., $\mu_0 = 0$ in Eq.~\ref{craigT};
and that the SNPs $i$ are completely independent such that we can
write the variance of the mean as the mean variance, ie,
${\var{\mean{D_i}} = \var{D_i}/s}$ in the denominator of Eq.~\ref{craigT}.
Below, we consider sources of deviation from $T \sim N(0,1)$ under the
null hypothesis.

\subsubsection{Deviations from $\mu_0=0$\label{anal1}}
In the large-sample limit, under the null hypothesis,
\begin{equation}
\lim_{n_F \rightarrow \infty} f_i =  p_i \; ; \; \; \lim_{n_G \rightarrow \infty} g_i = p_i  \, ,
\end{equation}
and hence 
\begin{equation}
\lim_{n_F, n_G \rightarrow \infty} D_i = \lim_{n_F, n_G \rightarrow \infty} \bigl(\abs{y_i - f_i} - \abs{y_i - g_i} \bigr) =0 \; .
\end{equation}
Intuition might further suggest that since $f_i$ and $g_i$ are both drawn
from binomial distributions which are symmetric about $p_i$, any
sampling deviations resulting from finite $n_F, n_G$ will fall
symmetrically, and hence $\mu_0 = 0$.   As we will show below, 
however, this conclusion is sensitive to two assumptions:
\begin{enumerate}
\item that the MAF differences between samples $F$ and $G$, $f_i - g_i$ are small;
\item that the sample sizes $n_F$ and $n_G$ are not only large, but comparable.
\end{enumerate}
Because the number of SNPs $s$ is quite large, slight deviations
away from $\mu_0=0$ have the power to shift the location of the null
distribution of $T$ considerably, rendering $T$ incomparable to
a standard normal unless the true
$\mu_0$ is known.  Consider that the difference in $T$ with and
without the $\mu_0 = 0$ assumption is
\begin{equation}
T - T_{\mu_0=0} = \frac{\mu_0}{\sqrt{\var{D_i}/s}} \,
\end{equation}
and that because $D_i$ ranges on $(-1,1)$, $\max (\var{D_i}) = 2$.  This means
that    
\begin{equation}
\min ( T - T_{\mu_0=0} ) = \frac{\sqrt{s}}{\sqrt{2}} \mu_0 
\end{equation}
which can be quite large for even small values of $\mu_0$ since the number of SNPs $s$ is on the order of $10^5$.  It is thus essential that $\mu_0$ be known or controllable.

\vspace{0.5em}\noindent{\textbf{Dependence of $\mu_0$ on slight differences in MAFs $f_i - g_i$.}}

Let us begin by writing the difference between MAFs $f_i$ and $g_i$
at locus $i$ as $\tau_i$,
\begin{equation}
f_i = g_i + \tau_i \, .
\end{equation}
We can then write
\begin{equation}
D_i = \abs{y_i - g_i - \tau_i} - \abs{y_i - g_i} \, ,
\end{equation}
and thus
\begin{align}
\mu_0 &= \mean{\abs{y_i - g_i - \tau_i} - \abs{y_i - g_i} } \, , \label{mu01}
\end{align}
where $\mu_0$ is $\mean{D_i}$ under the null hypothesis.

We next make a simplifying assumption:  since $p_i$ are the \textit{minor} allele frequencies and thus $0 \leq p_i \leq 0.5$, and since $f_i$ and $g_i$ are estimates of $p_i$, with few exceptions we will have $0 \leq f_i \leq 0.5$ and $0 \leq g_i \leq 0.5$ (eliminating this assumption does not significantly alter the results).  Under this assumption we can write
\begin{equation}
\abs{y_i - g_i - \tau_i} - \abs{y_i - g_i} =
\begin{cases}
\tau_i &\text{for $y_i=0$;}\\
-\tau_i &\text{for $y_i=0.5$;}\\
-\tau_i &\text{for $y_i=1$.}
\end{cases}
\end{equation}
and hence Eq.~\ref{mu01} may be written
\begin{align}
\mu_0 &= \sum_i \Bigl[ \tau_i \cdot \prob{y_i=0|p_i} - \tau_i \cdot \prob{y_i=0.5|p_i} - \tau_i \cdot \prob{y_i=1|p_i} \Bigr] \; \prob{p_i} \; \prob{\tau_i} \, ,
\label{mu03}
\end{align}
where $\prob{\cdot}$ denotes probability and where we have exploited the fact that because $F$, $G$ are independent samples of $P$, $\tau_i$ is independent of $p_i$, i.e., $\prob{\tau_i|p_i} = \prob{\tau_i}$.  Observing that 
\begin{equation}
\begin{aligned}
& \prob{y_i=0|p_i} = (1-p_i)^2 \, ; \\
& \prob{y_i=0.5|p_i} = 2p_i(1-p_i)\, ; \\
& \prob{y_i=1|p_i} =p_i^2 \, ,
\end{aligned} 
\end{equation}
Eq.~\ref{mu03} becomes
\begin{align}
\mu_0 &= \sum_i \bigl( 1-4p_i +2p_i^2 \bigr) \, \tau_i \; \prob{p_i} \; \prob{\tau_i} \label{mu04} \\
&= \mean{( 1-4p_i +2p_i^2 ) \, \tau_i}\, , \label{mu05} 
\end{align}
which is readily verified by simulation.

Eq.~\ref{mu05} implies that when $\tau_i$ deviates from zero, either
due to systematic differences in $F$ and $G$ (i.e., violation of
the assumption that both are drawn on the same population $P$) or
due to sampling variation, the location of the null distribution of the test statistic given by Eq.~\ref{craigT}
will be shifted by an amount equal to ${\mean{(1-4p_i +2p_i^2)\tau_i}\cdot\sqrt{s/\var{D_i}}}$ relative to that under the
assumption that $\mu_0 = 0$.  It is important to note that the
shift is a weighted average of $\tau_i$; ie, it
depends not only on the differences in MAFs $\tau_i$ but
also on $p_i$, 
and hence it is not sufficient that
$\mean{\tau_i}=0$, since 
small $\tau_i$ will be amplified when $p_i$ is small and reduced
when $p_i$ is large.  As a result, predicting the deviation away from
$\mu_0=0$ to properly calibrate $T$ requires knowing not only $\tau_i
= f_i - g_i$, but $p_i$ as well.

In practice, $\tau_i$ is easily calculated (examples of the distribution
of $\tau_i$ for the CGEMS and HapMap CEPH groups are given in
Fig.~\ref{freqdists}).  On the other hand, knowing $p_i$ requires making
assumptions about the population from which $Y$ is drawn.  In the case
where $Y$ is, in fact, drawn from a different underlying population
than are $F$ and $G$, the $p_i$ are difficult to obtain from the given
data and the shift in $T$ resulting from Eq.~\ref{mu05} is not readily
calculated.  (This effect is revealed in the empirical tests shown in 
Fig.~\ref{nulldistiswrong}, discussed in the empirical results 
section~\ref{res1} below, wherein the HapMap samples are shifted by
differing amounts.)



\vspace{0.5em} \noindent{\textbf{Dependence of $\mu_0$ on sample sizes $n_F$ and $n_G$.}}

The effect of deviations from the second assumption above is intuitively obvious: if $n_G > n_F$,
$G$ will better approximate the underlying population $P$ and so
will be closer on average to a future sample $Y$. 
The dependence is derived explicitly in the Appendix.

We can demonstrate this effect by simulation, as shown in
Fig.~\ref{AppFig1}.  Here, we begin by creating $10^5$ SNP MAFs
$p_i$ uniformly distributed on the interval $(0, 0.5)$.  From these
$p_i$, we simulate the $g_i$ with sample size $n_G=1000$ as given
by Eq.~\ref{idealg} (i.e., a binomial sample) as well as 200
independent samples $Y$ with $y_i$ as given by Eq.~\ref{idealy}.  By simulating
$f_i$ per  Eq.~\ref{idealf} as $n_F$ is varied and computing
$\mean{D_i}$ for each sample $Y$ per Eq.~\ref{craigD}, we can observe
the dependence of $\mean{D_i}$ under the null hypothesis (i.e.
$\mu_0$) on the sample size of $n_F$.  A plot of the result is
provided in Fig.~\ref{AppFig1}.  
As seen in the plot and derived
explicitly in the Appendix, the dependence in this case varies indirectly with 
${n_F}$; as expected based on the intuition above, smaller $n_F$
leads to larger values of $\mean{D_i}$, indicating that $Y$ is closer
to $G$ (the larger, more representative sample of $P$) than it is to $F$. 
Although the difference is small,
$\mean{D_i}/\sqrt{\var{D_i}/s}$ -- given in Fig.~\ref{AppFig1}(B)  -- is
quite large, which would lead to a high false-positive rate in
practice if the $\mu_0=0$ assumption were used and $T$ values compared to the
presumed null distribution $N(0,1)$.  Thus, we see that as $n_F$ decreases,
the distribution of $T$ under the null hypothesis diverges from the 
standard normal distribution, resulting in a higher false positive rate than
that predicted by the nominal $\alpha$ from the standard normal.

\subsubsection{Deviations from $\var{\mean{D_i}}=\var{D_i}/s$\label{anal2}}

Invocation of the central limit theorem to compare $T$ 
to a standard normal distribution (as given in
Eq.~\ref{craigT}) requires that the variance
of the mean of $D_i$ be estimable by the mean of the variance, ie,
${\var{\mean{D_i}}=\var{D_i}/s}$.  This, in turn, requires that the
$D_i$ are uncorrelated. However, if the various $D_i$ are correlated---most notably due to linkage disequilibrium---this is no
longer true.  Specifically, the variance of the mean for $s$ variables
$D_i$
with variance $\var{D_i}$ and average correlation $\rho$ 
amongst the distinct $D_i$ is given by
\begin{equation} 
\var{\mean{D_i}} = \left( \frac{1}{s} + \frac{s-1}{s} \rho \right) \var{D_i} \, .
\label{var}
\end{equation}
In the case where the average correlation amongst the $D_i$'s is
zero, Eq.~\ref{var} yields the result which is found in the denominator
of Eq.~\ref{craigT}; on the other hand, $\rho \neq 0$ generates a
$\bigl( 1+(s-1)\rho \bigr)$ multiplicative increase over the correlationless
variance.  The large number of SNPs $s$ results in little room for
any correlation between them: consider that Eq.~\ref{var} dictates that 
for a modest number of
SNPs $s=5\cdot10^4$ even a very slight average correlation between all
pairs of SNPs $\rho = 0.002$ would result in a tenfold 
increase in $\var{T}$; for 500K SNPs ($s=5\cdot10^5$), $\rho =
0.0002$ causes a a two order of magnitude increase in $\var{T}$.
However, it is impossible to ascertain $\rho$
simply from $y_i$, $f_i$, and $g_i$.  Instead, this issue may be addressed
by choosing fewer SNPs and assuming that $\rho$ is sufficiently small.

\subsection{Results of Empirical Tests\label{res}}
To demonstrate the results derived in Sect.~\ref{anal} above, as well
as to explore the performance of the method in realistic
situations, 
we carried out the computations described by Eqs.~\ref{craigD},\ref{craigT}
for various $F$, $G$, and $Y$ as described in  Table~\ref{testtab}.
Distributions of $T$  for each of the
17 tests described in Table~\ref{testtab} are shown in the corresponding 
figures listed in the table.
Bearing in mind the fact that $\abs{T}>1.64$ yields a  nominal $\alpha$
($p$-value) of 0.05
and $\abs{T}>4.75$ yields a nominal $\alpha=10^{-6}$ when compared to
a standard normal distribution, 
the vast majority of samples we tested which were in neither $F$
nor $G$ were misclassified as being members of one or the other group when
using the $\alpha=0.05$ threshold for rejection of the null hypothesis;
the misclassification rate was also higher than expected when
using a nominal $\alpha=10^{-6}$ threshold.
The high false-positive rate in practice is
attributable to sensitivity to the assumptions which underlie the method,
as described above in Sect.~\ref{anal}.  We present the results 
under the assumptions from~\cite{HOME08} and then discuss the possibility 
of improving them based on our analytical and empirical findings.

\subsubsection{Deviation from putative null distribution\label{res1}}

\noindent{\textbf{Choice of $F$ and $G$.}} 

In Sect.~\ref{anal1}, we saw that $T$ will depend on the characteristics
of the samples $F$ and $G$.
The effect is demonstrated in the results shown in Fig.~\ref{plotdists}.
In these plots,  $T$ statistics (Eq.~\ref{craigD},~\ref{craigT})  
are given for all the CGEMS and S.1--S.5 samples for three choices of $F$ and $G$:
\begin{itemize}
\item $F$ = HapMap CEPH, $G$ = CGEMS case;
\item $F$ = HapMap CEPH, $G$ = CGEMS control;
\item $F$ = CGEMS control, $G$ = CGEMS case.
\end{itemize} 
The distribution of minor allele frequencies for each of these
three groups (CGEMS cases, controls, and HapMap CEPHs) and the distribution of
MAF differences for all three pairs of these groups may be seen in
Fig~\ref{freqdists}.  Notably, even though it may reasonably be expected that
the HapMap CEPH sample closely resembles the Caucasian subjects in
CGEMS, the distributions of the allele frequencies is much more similar in
CGEMS cases and CGEMS controls than in either group and HapMap CEPHs. (The
most striking difference in the HapMap and CGEMS distributions occurs
around $0.5$, where it can be seen that the minor (MAF$<0.5$) allele in
the CGEMS samples sometimes has a frequency $>0.5$ in HapMap CEPHs.)
Importantly, the width of the the distribution of MAF differences $\tau_i = f_i-g_i$ is much greater
when HapMap CEPHs are one of $F/G$: although the mean difference
in allele frequencies is quite small (0.0003--0.001) in all cases,
$\var{\tau_i}$ is an order of magnitude larger when HapMap CEPH is used as one
of the the groups, leading to non-zero $\mu_0$ via Eq.~\ref{mu05}.
Additionally, the sample size of the HapMap group is much smaller than
that of CGEMS, thus biasing classification of an unknown sample toward 
the larger (and hence more representative) CGEMS sample when HapMap 
is used for one of the groups 
(cf. Sect.~\ref{anal1} and Appendix for associated derivations).

As expected, using the HapMap CEPHs for $F$ fails to separate the CGEMS
case and control distributions, such that CGEMS controls and cases all
yield high $T$ (and hence would all be classified as cases) when $G$ = CGEMS cases;
the situation is analogous for $G$ = CGEMS controls (Fig.~\ref{plotdists}, top and
center left).  Only in the situation where $F$ and $G$ have
similar large sample sizes and similar MAFs (when $G$ = CGEMS cases and $F$ = CGEMS controls) is good separation achieved,
with the $T$ statistics generally falling on the appropriate side of 0
(Fig.~\ref{plotdists}, bottom left); even so, 15 of the controls were
misclassified as cases.  This final case, which achieves 99.4\% accuracy
using $\abs{T}>1.64$ (nominal $\alpha=0.05$),
is analogous to the data presented in~\cite{HOME08},
for which all samples are in either $F$ or $G$.  As anticipated, the 
accuracy of the classification of cases and controls is dependent on the
choice of $F$ and $G$.

The classification of the 1600 samples described in Sect.~\ref{sim1} with
the same choices of $F$ and $G$ (right column of Fig.~\ref{plotdists})
is also instructive.  In all three cases, all samples achieve high
$T$ statistics despite the fact that they are in neither $F$ nor
$G$, frequently with $\abs{T}\gg 4.75$, i.e., a nominal $p$-value less
than $10^{-6}$. 
(No simulated sample genotype was identical to any true positive genotype at 
greater than 62\% of loci, comparable to the degree of genetic identity observed in the
real samples.)
That is to say, the method classifies as positive individuals who possess a genotype
$y_i$ that is drawn on $f_i$ or $g_i$, but who are not necessarily in $F$ or $G$.  This is unsurprising, since Eqs.~\ref{craigD},\ref{craigT} quantify the degree to which $Y$ is not equidistant from $F$ and $G$.  Furthermore, this suggests that relatives of true positives may be misclassified (we consider this below in Sect.~\ref{res3}).


%

\vspace{0.5em} \noindent\textbf{Classification of null samples when $F$ and $G$ are well-chosen.}

Having observed the sensitivity of the classifier to the appropriate
choice of $F$ and $G$, 
we now explore the classification of samples which are in neither $F$ nor $G$
in the case where $F$ and $G$ are well-chosen.  Here, we randomly select
100 cases and 100 controls from CGEMS to form an out-of-pool test sample
set comprising 200 individuals, and recompute the MAFs for the remaining
1045 CGEMS cases ($G$) and 1042 CGEMS controls ($F$).  (Several such random
subsets were created; the results were consistent and hence we present
a single representative one.)  SNPs were kept
subject to the same constraint (MAF$>0.05$ in both $F$ and $G$) as above,
and  $T$ statistics (Eq.~\ref{craigD},~\ref{craigT}) were computed for all
the test samples using $f_i$ and $g_i$ as described.

For the  positives samples (those in $F$ or $G$), the classifier
performs fairly well, correctly classifying 2083 samples (and calling
4 as in neither $F$ nor $G$).  However, of the 200 test samples which
were in neither $F$ nor $G$, only 62 have $\abs{T}<1.64$, and
the bulk are misclassified into the reduced group of CGEMS cases.
The rate of false positives is thus 69\% if $T$ is used as an
indicator of group membership under the assumptions in~\cite{HOME08}
at the nominal $\alpha=0.05$ 
(see Table~\ref{restab}).  A plot of the $T$ values for all samples is
given in Fig.~\ref{nulldistiswrong}(A).  A similar test, in which HapMap
individuals unrelated to the CGEMS participants (90 each from CEPH and
YRI groups) were classified against the same subsets of 1045 CGEMS cases
($G$) and 1042 CGEMS controls ($F$), yields similar results: all the
YRI individuals and 85/90 of the CEPH individuals were  misclassified
into the group of CGEMS cases at $\alpha=0.05$; a plot of the $T$ value
distributions are given in Fig.~\ref{nulldistiswrong}(B).  Selecting a
more stringent $\alpha=10^{-6}$ (the minimum reported in~\cite{HOME08})
results in a 29.5\% false-positive rate amongst the 200 out-of-pool
CGEMS samples, 72\% false-positive rate amongst HapMap CEPHs, and 100\%
false-positive rate amongst HapMap YRIs.  A summary of the specificity
and sensitivities obtained in this test is given in Table~\ref{restab}.

The reason for the high false-positive rates in practice despite the
stringent nominal false positive rate is clear from the plots
Fig.~\ref{nulldistiswrong}(A,B): namely, it can be seen that the putative null
distribution (light grey line, $N(0,1)$, cf Eq.~\ref{craigT}) does
not correspond to the observed distribution for samples for which
the null hypothesis is correct, with differences in both the location
and width. 

The overall shift to the right is a product of the small differences in
$f_i - g_i$ which accumulate as given by Eq.~\ref{mu05}.  Because in
this test we happen to know the MAFs $p_i$ along with $f_i$ and $g_i$ for each
of the CGEMS samples, we can compute $\mu_0$ given by Eq.~\ref{mu05}
as $1.133\cdot10^{-4}$ and verify that, when divided by the average
$\sqrt{\var{D_i}/s}\approx 5.6\cdot10^{-5}$ amongst the samples,
the center of the observed null distribution will be at $T\approx 2$.  Indeed,
visual inspection of Fig.~\ref{nulldistiswrong}(A) shows that shifting
each $T$ distribution by -2 would result in $F$, $G$, and null-sample
distributions which lie more symmetrically about $T=0$.   Note, also, that the 
HapMap CEPHs and YRIs are shifted by different amounts than are the CGEMS
samples, due to the fact that the $p_i$'s which underlie the HapMap samples
differ from each other and from CGEMS.  From this, we can see that samples
$Y$ which are not drawn on the same population as $F$ and $G$ may in practice have a high false positive rate.

The effect of LD, derived in Sect.~\ref{anal2}, is also seen in these
examples.  In Fig.~\ref{nulldistiswrong}(B), we observe a narrower
distribution of $T$ for the HapMap YRI samples versus the Caucasian CGEMS
participants and HapMap CEPHs (the Yoruban individuals, who come from an
older population, have lower average LD).  The same effect is observed by comparing the
distribution of $T$ for the simulated samples in Fig.~\ref{plotdists}
(for which each SNP was independently sampled and hence have artificially
low LD) to those of real populations.  


\subsubsection{Correcting for deviations from $N(0,1)$\label{res2}}

Although the empirical false-positive rates obtained the the tests
described above are exceedingly high, the distributions of $T$
obtained in Fig.~\ref{nulldistiswrong}(A,B) are nonoverlapping.
Hence, one might expect that if one could appropriately calibrate
the thresholds of $T$ at which classification is made, the 
sensitivity and specificity of the test could be considerably improved.
(Note that, in practice, one does not know where the true-positive $F$ and
$G$ distributions of $T$ lie; this requires the genotypes of the $F$ and
$G$ individuals.)  Two approaches may be taken toward calibrating 
classification thresholds for $T$: an analytical approach, based on the
results in Sect.~\ref{anal} above; or an empirical approach, based on
constructing a null distribution from available samples.  As we will
see, both these approaches pose substantial difficulties.

\vspace{0.5em} \noindent{\textbf{Analytical approach.}}

In order to correct for the deviations from $N(0,1)$ analytically,
we need to know both the location and width of the distribution of $T$
in the non-ideal circumstances under which the test is being conducted.
That is, we need to know deviations from $\mu_0=0$ resulting from MAF
differences $f_i - g_i$ and sample size differences of $n_F$ and $n_G$
(cf. Sect.~\ref{anal1} and Appendix), as well as the average correlation
amongst SNPs $\rho$ (cf. Sect.~\ref{anal2}, Eq.~\ref{var}).  

Let us first consider the result in Eq.~\ref{mu05}, which shows that
$\mu_0$ in practice will be a function of the MAF differences $\tau_i =
f_i - g_i$ as well as the MAFs $p_i$ of the population $P$ of which $Y$
is a sample.  
If we are well-assured that $F$ and $G$ are
large samples of the same population $P$ and that $Y$ is also a sample
of that population, an average of $f_i$ and $g_i$ may be used to
estimate $p_i$ (the $y_i$, while necessarily drawn on $p_i$, are
too small a sample to be a good estimate) and thus obtain $\mu_0$.
Results of this approach (for the tests shown in Fig.~\ref{nulldistiswrong}
and Table~\ref{restab})
are given in Table~\ref{restab2}, in which $p_i$ was estimated as $(n_G\cdot g_i + n_F \cdot f_i)/(n_G+n_F)$ and $\mu_0$ was computed according to Eq.~\ref{mu05}.  A slight improvement in
the performance of the method can be seen by comparing the first
two columns of Table~\ref{restab} to those of Table~\ref{restab2}.

However, the assumption used to compute $p_i$ (i.e., that $Y$,
$F$, and $G$ are all i.i.d. samples of the same population $P$) is
one on which the accuracy of the correction is strongly dependent;
consider, for instance, that the $\mu_0 \approx 1.133\cdot10^{-4}$
obtained for the simulations in Fig.~\ref{nulldistiswrong}(A,B) and
discussed above will produce the appropriate shift $T \approx T_{\mu_0=0}
- 2$ for the 200 CGEMS samples in Fig.~\ref{nulldistiswrong}(A) using
this method, but will not centralize the HapMap $T$ distributions in
Fig.~\ref{nulldistiswrong}(B) appropriately, because the $f_i$ and $g_i$
are not good estimates of the MAFs of the populations from which
the HapMap samples are drawn.  Applying this correction to the HapMap
samples (equivalent to moving the HapMap $T$ distribution two units to
the left in Fig.~\ref{nulldistiswrong}(B)) results in a misclassification
rate of 86\% (nominal $\alpha=0.05$) and 44\% (nominal $\alpha=10^{-6}$)
for the HapMap CEPHs and continued 100\% misclassification of all HapMap
YRIs.  It is thus essential that if the $\mu_0$ given by Eq.~\ref{mu05}
is to be used, sound estimates of $p_i$ need to be obtained.  When $Y$
is not a sample of the same population as $F$ or $G$, estimates of $p_i$
are unobtainable from $f_i$, $g_i$ and $y_i$ alone, and
hence this correction relies upon the assumption that $G$, $F$, and $Y$ are
well-matched.

The second influence on $\mu_0$, described in both Sect.~\ref{anal1} and
the Appendix, is the effect of the sample sizes $n_F$ and $n_G$.  Here,
corrections are readily made, provided the sample sizes of $F$ and $G$
are known.  In a forensics context, where $G$ is a sample of unknown
composition, $n_G$ may not be known; on the other hand, in other contexts
(such as when using case and control MAFs from a GWAS), sample sizes are
known and readily adjusted for.   (In this test, $n_F \approx n_G \approx 1000$,
and the correction is negligible.)

We also saw in Sect.~\ref{anal2} and Fig.~\ref{nulldistiswrong}(B)
that the distribution of $T$ for null samples will depend on the degree
of correlation between the SNPs.  To accurately  derive the width of
the $T$ distribution for null samples, one would need to either select
SNPs that yield vanishingly small $\rho$ or know the value of $\rho$
with high accuracy for the population of which $Y$ is a sample so
that it can be discounted.  The latter option requires knowledge beyond
the MAFs of $F$ and $G$ and the genotype of individual $Y$; namely, it
requires multiple genotypes from the population $P$ from which $Y$
was drawn such that the average correlation $\rho$ between SNPs can be
computed; even with a collection of null genotypes, the computation of
the average pairwise correlation for $10^5$ SNPs is a computationally
unfeasible task.  Rather, selecting fewer SNPs in order to reduce LD
is a more workable solution; the results of this approach can be seen
in Fig.~\ref{nulldistiswrong}(C,D) and in Table~\ref{restab}.  Here,
50,000 SNPs were selected, uniformly distributed across of the $481,382$ 
SNPs used in Fig.~\ref{nulldistiswrong}(A,B).  50,000 SNPs
was shown in~\cite{HOME08} to be a reasonable lower bound to detect at
nominal $\alpha\approx10^{-5}$
one individual amongst 1000, which is 
the concentration of true positive individuals in this test.

As is clear from Fig.~\ref{nulldistiswrong}, reducing the number of
SNPs narrows the distributions considerably, yet at the same time brings
them closer together such that the crisp separation previously obtained
is reduced.  Using this method, we see that the 200 CGEMS samples now
have a distribution closer to that of the putative null $N(0,1)$ such
that using a threshold of $\alpha=0.05$ yields an improved---yet still
larger than nominal---21\% false-positive rate while maintaining
a high 96.3\% true positive rate.  However, the misclassification rate is
still over 50\% for both HapMap samples, and improving these values
requires compromising the sensitivity, a direct result of the overlapping
$T$ distributions for the $G$ and HapMap samples.

Finally, we can consider applying both the SNP reduction and the $\mu_0$
correction applied above; the results here are given in the final two 
columns of Table~\ref{restab2}.  Because $F$ and $G$ are well-matched and
the $\mu_0$ correction given by Eq.~\ref{mu05} is slight in the case of
these 50,000 SNPs, the correction happens to offer little improvement over that
achieved by subsetting the SNPs.

\vspace{0.5em} \noindent{\textbf{Empirical approach.}}

Another potential approach to obtaining a correct null distribution is
purely empirical, 
namely, collecting a set of presumed-null genotypes (called $N$)
which can be assumed to be drawn from the same population as $Y$, and
determining the distribution of $T$ for the null samples $N$.
However, once again the method's sensitivity to the assumptions 
are a source of error.

To see this, let us once more return to Fig.~\ref{nulldistiswrong}. 
In these figures, vertical bars represent the 0.05 and 0.95 quantiles
of the 200 CGEMS (black), 90 HapMap CEPH (cyan) and 90 HapMap YRI (blue)
$T$ distributions.

Let us first consider a situation in which we have $f_i$ and $g_i$, along
with an individual $Y$ who is one of the 200 CGEMS samples not in $F$
or $G$,  but no other genotypes.  We might reasonably turn to publicly
available HapMap genotypes as our group $N$ from which we construct an
empirical null distribution from which we set thresholds.  The lines in Fig.~\ref{nulldistiswrong}(A,C)
depict this case.  Using thresholds obtained from the HapMap CEPH
distribution (cyan lines) still incorrectly classifies half of the
200 CGEMS samples; the false positive rate is yet greater (and the
true-positive rate smaller) when using the HapMap YRI distribution.
These lines illustrate the importance of choosing for $N$ a sample which
closely resembles $Y$---as with the choice of $F$ for a given $G$ in
Sect.~\ref{res1}, HapMap CEPHs are insufficiently similar to CGEMS to provide
accurate results, despite the fact that both samples are Americans
of European descent.  

The converse is true as well: if we have $N$, $F$, and $G$ which are
well matched---such as illustrated in Fig.~\ref{nulldistiswrong}(B,D),
in which $N$, $F$, and $G$ all come from CGEMS data---yet $Y$ is not
drawn from the same underlying population as $N/F/G$, the method will
incorrectly classify $Y$; roughly a quarter of the HapMap CEPHs and
the majority of HapMap YRIs lie outside the thresholds set by  the 200
CGEMS samples in Fig.~\ref{nulldistiswrong}(B,D).  Once again, this
underscores the importance of the assumption that $F$, $G$, and $Y$ are
all i.i.d. samples of the same population $P$, and---if a sample $N$
is being used to construct a null distribution empirically---it, too,
must be an i.i.d. sample of $P$.

Another empirical option is that of simulating genotypes from the $f_i$
and $g_i$ to simulate $T$ under the \textit{alternative} hypothesis, with
the assumption that the null and alternative hypothesis $T$ distributions
do not strongly overlap.  However, this method also requires that  $F$ and
$G$ are large and well-matched samples, since (as can be seen in the top-
and middle-right graphs in Fig.~\ref{plotdists}) poorly-matched $F$ and
$G$ will not produce crisply separated distributions.
Furthermore, the thresholds derived by this approach will relate not to
the false-positive rate but rather to the false-negative rate, i.e., these
thresholds would control the power of the test, and the specificity
in practice will remain unknown.

We have thus seen that small deviations from the assumptions
that $F$, $G$, and $Y$ are i.i.d. samples of the same population $P$
can produce false-positive rates which greatly exceed those predicted
by the null hypothesis.  Even when these sources of error were adjusted
for, in our tests we still observed a false positive rate that
was higher than expected, such that the false positive rate was never
less than 20\% in practice for a nominal false-positive rate of 5\%,
and never less than 13\% at a nominal false-positive rate of 0.0001\%.
While the distributions of $T$ for the $F$, $G$, and various $Y$ samples
were observed to be separate in Fig.~\ref{nulldistiswrong}, we find that
calibrating
the thresholds accurately in absence of genotype information for $F$
and $G$ is not obviously doable.  More importantly, it is not clear that,
once thresholds are chosen, the empirical specificity could be assessed
without additional genotype information from subjects who are well-matched
to $F$, $G$ and $Y$.

\subsubsection{Positive predictive value of the method.}
The effect of the modest specificity---even in the best of cases described
above---on the posterior probability that the individual $Y$ is in $F$
or $G$ is considerable, given that the prior probability is likely to be
relatively small in most applications of this method.  Let us
consider the positive predictive value (PPV), which quantifies the post-test
probability that an individual $Y$ with a positive result (i.e., significant $T$) is
in $F$ or $G$.  This
probability depends on the prior probability that the individual is
in $F$ or $G$, i.e., on the prevalence of being a member of $F$ or $G$.
PPV follows directly from Bayes' theorem, and is defined as
\begin{equation}
\mathsf{PPV} = \frac{\mathsf{Sens}\cdot\mathsf{Prev}}{\mathsf{Sens}\cdot\mathsf{Prev}+(1-\mathsf{Spec})(1-\mathsf{Prev})} \, ,
\label{ppv}
\end{equation}
where the PPV is the posterior probability that $Y$ is in 
$F/G$ given a prior probability of $\mathsf{Prev}$.  We can write this 
equivalently in terms of the positive likelihood ratio $\mathsf{LR}_+$,
\begin{align}
\mathsf{Posterior\, odds} &= \mathsf{LR}_+ \cdot \mathsf{Prior\, odds}\\
 \mathsf{LR}_+ &=  \frac{\mathsf{Sens}}{(1-\mathsf{Spec})}  \,.
\label{posto}
\end{align}
A plot of PPV vs. prevalence is given in Fig.~\ref{ppvplot}.  Even with
the best sensitivity (99.23\%) and specificity (87\%) obtained in
our tests---that in which $F$, $G$, and $Y$ were drawn on the same
underlying population $P$, $\mu_0$ was accurately computed, and a nominal
$\alpha=10^{-6}$ was used as a threshold (cf. Table~\ref{restab2})---the
prior probability (prevalence) of $Y$ being in $F/G$ needs to exceed 54\%
in order to achieve a 90\% post-test probability that the subject is in
$F/G$.  For a PPV of 99\%, the prior probability needs to exceed 72\%
for any specificity under 95\%, assuming the observed sensitivity of 99\%.
We thus see the strong need for prior belief that $Y$ is in $F$ or $G$.

The difficulty in assessing the (empirical) specificity of
the test in absence of additional data makes the posterior probability
difficult to ascertain since the false positive rate in practice is much
greater than that given by the nominal false-positive rate $\alpha$.
Eq.~\ref{posto} underscores this fact; referring once more to the best
result in Tables~\ref{restab},~\ref{restab2}, consider that $LR_+$
at 87\% specificity and 99\% sensitivity is 7.6, versus  990000 if the
nominal false-positive rate $\alpha=10^{-6}$ were correct.  For prior
probability of 1/1000, the first case yields a posterior probability of
1.1/1000, while the second yields a posterior probability of 998/1000.
These differences, which are difficult to measure without additional,
well-matched null sample genotypes and which depend strongly on the
degree to which the assumptions underlying the method are met (consider
the differences between the CGEMS and HapMap CEPH specificities in
Tables~\ref{restab},~\ref{restab2}), pose a severe limitation on
the utility of using Eqs.~\ref{craigD},\ref{craigT} to resolve $Y$'s
membership in samples $F$ or $G$.


\subsubsection{Classification of relatives\label{res3}}
We now turn to the classification of individuals who are relatives
of true positives.  As discussed above in Sect.~\ref{res1}, 
the results from simulations S.1--S.5 in Fig.~\ref{plotdists} suggest
that individuals who are genetically similar, but not identical to,
the subjects in pools $F$ and $G$, frequently exhibit high $\abs{T}$.
This effect can be investigated by using HapMap families, since we
can reasonably expect that the children will bear a greater resemblance
to their parents than their parents do to one another.  Recalling that
the HapMap pools consist of thirty individual mother-father-offspring
pedigrees, we construct pools as follows:
\begin{itemize}
\item $F$ = Mothers from pedigrees 1--15 and fathers from pedigrees 1--15
\item $G$ = Children from pedigrees 1--15 and fathers from pedigrees 16--30
\end{itemize}
and then compute $T$ for mothers and children from pedigrees 16--30 using
the same SNP criteria as before.
The results of these tests for both the
CEPH and YRI pedigrees, given in Fig.~\ref{hapclass2}, are as expected,
with the children having a significantly higher distribution of $T$
than the mothers; the $T$ values for all the children were so large that
$p$-values $\ll 10^{-16}$ were obtained when comparing to $N(0,1)$.
By contrast, 5/15 of the YRI mothers from pedigrees 16--30 and
10/15 of the CEPH mothers from pedigrees 16--30 yielded $\abs{T}>1.64$ (with
distributions roughly centered about $T=0$).
The wider distribution amongst the CEPHS again reflects the effect
of LD. In Fig.~\ref{hapclass2} we can see that the method has
the power to resolve three groups: those in a group, those related
to members of a group, and those who are neither (as the
groups become bigger, and hence more homogeneous, we would expect the
distributions to move closer together, as evidenced by the lower range of
$T$ for the CGEMS-based tests in Fig.~\ref{plotdists}).  Note, however,
that without knowing the distribution of $T$ for true positives (which
necessitates knowing the genotypes of true positives) setting a threshold
to distinguish between true positives and their relatives is not possible
by any of the methods described above.

In order to explore the effect of genetic similarity in a controlled,
ideal situation for which $F$ and $G$ are known to be samples of the same
underlying population and for which all SNPs are known to be independent
(i.e., in the ideal situation in which the putative null distribution $N(0,1)$ should
hold), we carried out the simulations described in Sect.~\ref{sim2}. In
these simulations,  the underlying population $P$ was taken to have MAFs $p_i$
as given by the CGEMS controls;  $f_i$, $g_i$, and $y_i$ were
derived as described in Sect.~\ref{sim2} as binomial samples of $p_i$.

In the first of these simulations, the test samples were constrained to
have a proportion $q$ of SNPs identical to a true positive individual, with
the remaining SNPs drawn on $p_i$.  A plot of the false
positive rate, defined as the fraction of the 200 simulated samples
that achieve significant $\abs{T}>1.64$ ($\alpha<0.05$), as the similarity parameter $q$ is varied
is shown in  Fig.~\ref{misclass}.  Once simulated samples exceeded $65\%$
identity with a true positive individual, they universally achieve
significant $T$, and significant values of $T$ are found over half the time for
simulated samples exceeding  $60\%$ identity.  (It should be noted that
of the real samples, no two had $>62\%$ fractional identity.)

In the second set of these simulations, the test  samples were drawn
from a weighted mixture of MAFs:
\begin{eqnarray}
y_{i} &\sim& \mathsf{Bin}(2, p'_i) / 2 \, , \\
p'_i &=& (1-q) p_i + (q) g_i \, ,
\end{eqnarray}
i.e., the sample was drawn from MAFs $p'_i$ which are  $q$ percent like $G$ and
$(1-q)$ like CGEMS controls (MAFs $p_i$).  By simulating 200 samples for various $q$,
computing $T$ for each sample using the simulated $F$ and $G$, and
counting the number of samples that achieve significant $\abs{T}>1.64$ at $\alpha=0.05$, we can see how
the false positive rate varies with the percentage of $G$. Results are
given in Fig~\ref{misclass}.  The misclassification rate exceeds 50\%
for $q=0.05$; at $q=0.1$, all samples yield significant $T$.

The misclassification of relatives follows directly from the
method's premise.  Eqs.~\ref{craigD},\ref{craigT} together answer
whether \textit{individual's genotype $y_i$ is closer to sample $G$'s MAFs $g_i$ than to sample $F$'s MAFs $f_i$
than would be expected by chance}, and it is unsurprising that a
relative of a true member of $G$ would appear closer to $G$ (via
Eqs.~\ref{craigD},\ref{craigT}) than to $F$.  

Put another way, $Y$ being a member of $G$ is \textit{sufficient}
but not \textit{necessary} for $y_i$ to be closer (via Eq.~\ref{craigD})
to $g_i$ than to $f_i$; it is possible for other sources of genetic
variation to cause $y_i$ to be closer  $g_i$ than to $f_i$.  We can
observe this by turning once again to Fig.~\ref{nulldistiswrong}(A,C),
where the dashed red and green lines show that the not-in-$G$
CGEMS cases had a distribution of $T$ closer to the other CGEMS cases $G$,
and the not-in-$F$ CGEMS controls had a distribution of $T$ closer to the
other CGEMS controls $F$, indicating that small class-specific
genetic differences can yield altered values of $T$.  The erroneous
inferential leap that significant $T$ results from $Y$'s presence in $F$
or $G$ is responsible for the misclassification of relatives as well as
for misclassification of non-relatives in the previous examples.


\section{Discussion and Conclusions}

In this work, we have further characterized and tested the
genetic distance metric initially proposed in~\cite{HOME08}.  This
metric, summarized here by Eqs.~\ref{craigD},\ref{craigT},
quantifies the distance of an individual genotype $Y$ with respect to
two samples $F$ and $G$ using the marginal minor allele frequencies
$f_i$ and $g_i$ of the two samples and the genotype $y_i$.  The
article~\cite{HOME08} proposes to use this metric to infer the presence of
the individual in one of the two samples, and the authors demonstrate the
utility of their classifier on known positive samples (i.e., samples which
are in either $F$ or $G$) showing that in this situation their method
yields classifications of high sensitivity.  Our investigations reveal
that while the sensitivity is quite high (correctly classifying true
positives into groups $F$ and $G$) the specificity is considerably less
than that predicted by the quantiles of the putative null distribution
$N(0,1)$.  As a result, Eqs.~\ref{craigD},~\ref{craigT} are severely
limited in their utility for discerning  $Y$'s presence in samples $F$
or $G$.

In this work we have shown
that high $T$ values, significant when compared against
$N(0,1)$, may be obtained for samples that are in neither of the 
pools tested under several circumstances:
\begin{itemize}
\item when pools $F$ and $G$ are sufficiently dissimilar such that the
differences in $f_i$ and $g_i$ dominate, 
as seen in Sects.~\ref{anal1} and \ref{res1} as well as the Appendix;
\item when $Y$ is a sample of a different population than are $F$ and $G$,
as seen in Sect.~\ref{res1};
\item when a small amount of average LD is present such that the putative null distribution in Eq.~\ref{craigT} does not hold (due to a violation of the CLT assumption of independence), as seen in Sects.~\ref{anal2} and \ref{res1};
\item and when a sample is genetically similar, but not identical to, individuals 
comprising $F$ or $G$ (e.g., relatives of true positives),  as seen in Sect.~\ref{res3}.
\end{itemize}
The high false positive rates in the first two cases result from 
assumptions underlying the putative null distribution which are not
met in practice, specifically, that the individual $Y$ along with samples
$F$ and $G$ are all i.i.d. samples of the same underlying population $P$,
and that the amount of correlation between all $s$ SNPs is vanishingly 
small. As we saw in Sect.~\ref{res1} and \ref{res2}, these assumptions are difficult to meet;
for instance, HapMap CEPH and CGEMS samples are sufficiently
dissimilar that they introduce error in violation of the first assumption,
despite the fact that both samples are Americans of European descent.  
Adjusting for deviations from the putative null distribution also
requires making strong assumptions or obtaining additional information,
as seen in Sect.~\ref{res2}.

Additionally, the conclusion that high $T$ values result from $Y$'s
presence in $G$ relies upon the questionable assumption that individuals
in neither $F$ nor $G$ will be equidistant from both, resulting in false
positives even when the other assumptions are met.  For instance,
similarly genotyped individuals (both relatives and simulated samples)
are often classified into the same group despite the fact that the 
other assumptions were met (Sect.~\ref{res4}).  Amongst non-relatives,
even when the thresholds have been adjusted for
violations of the above assumptions as in Sect.~\ref{res2}, Eqs.~\ref{craigD},\ref{craigT}
produce misleading classifications  at a rate that is considerably greater
than expected (21\% vs. nominal 5\% and 13\% vs. nominal 0.0001\% in
the best cases reported in Table~\ref{restab}).  The unpredictable
false positive rate in practice, resulting from the difficulty in
accurately calibrating the significance of $T$, results in a likelihood
ratio (and hence post-test probability) that is also unpredictable,
with higher false positive rates yielding lower post-test probabilities.
When the prior probability of $Y$'s presence in $F$ or $G$ is modest,
strong evidence (i.e., high specificity) is needed to outweigh
this prior, which was not achieved in our tests (Sect~\ref{res3}).  On the other hand,
when samples were known \textit{a priori} to be in one of the groups
$F$/$G$, Eqs.~\ref{craigD},\ref{craigT} correctly identify the sample
of which the individual is part (Sect.~\ref{res3}).

These findings have implications both in forensics (for which the
method~\cite{HOME08} was proposed) and GWAS privacy (which has 
become a topic of considerable interest in light of~\cite{HOME08}).
We briefly consider each:

\vspace{0.5em}\noindent\textbf{Forensics implications.}

The stated purpose of the method---namely, to positively identify
the presence of a particular individual in a
mixed pool of genetic data of unknown size and composition---is
difficult to achieve.  In this scenario, we have $g_i$ (from
forensic evidence) and a suspect genotype $y_i$.  To apply the
method, we would need 1) to assume that $Y$ and $G$ are indeed
i.i.d. samples of the same population $P$; 2) to obtain a sample
$F$ which is \textit{also} a sample of the underlying population $P$,
well-matched in size and composition to $G$;
3) to obtain an estimate of the sample size of $G$ such that sample-size
effects can be appropriately discounted; and 4) to assume that the
$p$-values at the selected classification thresholds are accurate.
We have seen in the Results 
section the sensitivity to the assumption that $Y$, $F$,
and $G$ all come from the same population, the sensitivity to the sample
size of $G$, and the difficulties in calibrating thresholds; the high
false-positive rates which result from even small violations of these assumptions 
make it exceedingly likely that an innocent party will be wrongly
identified as suspicious; its is even more likely for a relative
of an individual whose DNA is present in $G$.

\vspace{0.5em} \noindent\textbf{GWAS privacy implications.}

Here the scenario of concern is that of a malefactor with the genotype
of one (or many) individuals, and access to the case and control MAFs
from published studies; could the malefactor use this method to discern
whether one of the genotypes in his possession belongs to a GWAS subject?
In this case, $F$ and $G$ are known to be samples of the same underlying
population $P$ (due to the careful matching in GWAS), and their sample
sizes are large and known.  However, the malefactor still needs 1)
to assure that $Y$ is a member of this population as well (as shown by
the poor results when HapMap samples were classified using CGEMS MAFs)
and 2) to assume that the $p$-values at the selected classification
thresholds are accurate.  Additionally, the prior probability that any
of the genotypes in the malefactor's possession comes from a GWAS subject
is likely to be quite small, since GWAS samples are a tiny fraction of
the population from which they are drawn.  Even if the malefactor
were able to narrow down the prior
probability to one in three, a sensitivity of 99\% and a specificity of
95\% is needed to obtain a 90\% posterior probability that the individual
is truly a participant.

On the other hand, if the malefactor \textit{does} have prior knowledge
that the individual $Y$ participated in a certain GWAS but does not know
$Y$'s case status, Eqs.~\ref{craigD},~\ref{craigT} permit the malefactor
to discover with high accuracy which group $Y$ was in.  Additionally,
in the case of \textit{a priori} knowledge, the participant's genotype
is not strictly necessary, since a relative's DNA will yield a large $T$
score that falls on the appropriate $F/G$ side of null.

Despite these limitations, we have found that the distance metric
(Eqs.~\ref{craigD},~\ref{craigT}) may still have forensic and
research utility. It is clear from both our studies and the original
paper~\cite{HOME08} that the sensitivity is quite high; in the (rare)
case that a sample has an insignificant $\abs{T}<1.64$, it is very
likely that $Y$ is in neither $F$ nor $G$.  We can also see that
genetically distinct groups have $T$ distributions with little overlap
(Fig.~\ref{nulldistiswrong}), and so it may be worth investigating the
utility of Eqs.~\ref{craigD},\ref{craigT} for  ancestry inference.

On this note, let us once more consider the quantity which
Eq.~\ref{craigD} measures, namely the distance of $y_i$ from
$f_i$ relative to the distance of $y_i$ from $g_i$.   Referring to
Fig.~\ref{plotdists} (right column) and Fig.~\ref{nulldistiswrong}(A,C),
we can see that samples $Y$ which are more like those in sample $G$ have
a distribution that lies to the right of samples which are more similar
to $F$, as expected; for example, in Fig.~\ref{nulldistiswrong}(A,C),
the distribution of null (not in $F,G$) CGEMS cases (dashed red line)
is shifted to the right with respect to the distribution of null CGEMS
controls, as might be expected from Eq.~\ref{craigD}, i.e., the CGEMS
case $Y$s are closer to CGEMS case $G$s than are the CGEMS control $Y$s.
Although this difference is not statistically significant, one could
imagine that it may be possible to select SNPs for which the shift
is significant, i.e., a selection of SNPs for which unknown cases
are statistically more likely to be closer (via Eq.~\ref{craigD}) to
the cases in $G$ and unknown controls are statistically more likely
to be closer to the controls in $F$.  In this case, a subset of
SNPs known to be associated with disease may potentially be used
with Eqs.~\ref{craigD},~\ref{craigT} to predict the case status of
new individuals; conversely, finding a subset of SNPs which produce
significant separations of the test samples may be indicative of a group
of SNPs which play a role in disease.  Because this type of application
would use fewer SNPs and would involve the comparison of two distributions
of $T$ (cases $\notin \{F,G\}$ vs. controls $\notin \{F,G\}$), it may
be possible to circumvent some of the problems stemming from the unknown
width and location of the null distribution described above; still, much
work is needed to investigate this possible application.  If successful,
the metric proposed in~\cite{HOME08}, while failing to function as a
tool to positively identify the presence of a specific individual's DNA
in a finite genetic sample, may if refined be a useful tool in the
analysis of GWAS data.


\renewcommand{\theequation}{A-\arabic{equation}}
\setcounter{equation}{0}
\section*{Appendix: Dependence of $\mu_0$ on the sample size of $F$ and $G$}

Consider  $\mean{D_i}$ (cf. Eq.~\ref{craigD}) under the null hypothesis assumptions that $Y$, $F$, and $G$ are all drawn i.i.d. from the same underlying population $P$ with MAFs $p_i$.  Writing the probability distribution of $p_i$ as $\prob{p_i}$, $\mean{D_i}$ is given by
\begin{align}
\mean{D_i} = &\mean{\abs{y_i - f_i} - \abs{y_i - g_i}} 
	=  \mean{\abs{y_i - f_i}} - \mean{\abs{y_i - g_i}}\\
\begin{split}
         = &\iiint_{-\infty}^{\infty} \abs{y_i - f_i} \; \prob{y_i | p_i}\, \prob{f_i | p_i} \, \prob{p_i}\, dy_i \, df_i \, dp_i - \\
         &\qquad \qquad- \iiint_{-\infty}^{\infty}  \abs{y_i - g_i} \; \prob{y_i | p_i}\, \prob{g_i | p_i} \, \prob{p_i}\, dy_i \, dg_i \, dp_i \; ,
\end{split}
\label{int1}
\end{align}
where we exploit the fact that $Y$, $F$ and $G$ are independent of each other but depend on the underlying population MAFs.

The dependence of the first (second) term in Eq.~\ref{int1} on $n_F$
($n_G$) is derived as follows.  First, we note that since each $y_i$ is
two Bernoulli trials (two alleles) with probability $p_i$, we have
the following values of $\abs{y_i - f_i}$ with probability $\prob{y_i | p_i}$
for each allowable value of $y_i$:
%
%
\begin{equation}
\abs{y_i - f_i} \cdot \prob{y_i | p_i} = 
\begin{cases}
\bigr( 1-f_i \bigl) \cdot \bigr( p_i^2 \bigl) &\text{for $y_i=1$}\, ;\\
\bigr( \abs{0.5-f_i} \bigl) \cdot \bigr( 2p_i(1-p_i)\bigl) &\text{for $y_i=0.5$}\, ;\\
\bigr( f_i \bigl) \cdot \bigr( (1-p_i)^2\bigl) &\text{for $y_i=0$}\, .
\end{cases}
\end{equation}
Moreover, since  each $f_i$ follows a binomial distribution of size $2n_F$
(two alleles per person),
we invoke the normal approximation to the binomial for 
values of $n_F>10$ with mean $p_i$ and variance $p_i(1-p_i)/(2 n_F)$. Hence:
\begin{eqnarray}
\prob{f_i | p_i} &=& \sqrt{\frac{2 n_F}{2 \pi p_i(1-p_i)}} \, \exp{\left[-\frac{2 n_F (f_i -p_i)^2}{2 p_i(1-p_i)}\right]} \\
&=& \frac{A_{F,i}}{\sqrt{\pi}} \exp{ \bigl[ -A_{F,i}^2 (f_i-p_i)^2 \bigr] } \, ,
\label{probf}
\end{eqnarray}
where we introduce 
\begin{align}
A_{F,i} = \sqrt{n_F/(p_i(1-p_i))}
\label{defna}
\end{align}
to simplify the notation.  In consequence, the first term of Eq.~\ref{int1} can be written:
\begin{multline}
\iint_{-\infty}^{\infty}  \biggl[ (1-f_i) (p_i^2) + (\abs{0.5-f_i}) (2p_i(1-p_i)) + (f_i) ((1-p_i)^2) \biggr] \cdot \\ \cdot \frac{A_{F,i}}{\sqrt{\pi}} \exp{ \biggl[ -A_{F,i}^2 (f_i-p_i)^2 \biggr] } \prob{p_i} \, df_i \, dp_i 
\label{int2}
\end{multline}
and the second term may be written analogously for $G$. The absolute value in Eq.~\ref{int2} is dealt with by considering the $f_i \geq 0.5$ and $f_i\leq 0.5$ cases separately, i.e., treating  Eq.~\ref{int2} as the sum of integrals
\begin{multline}
\int_{-\infty}^{\infty} \left[ \int_{0.5}^{\infty}  \biggl( (1-f_i) (p_i^2) + (f_i-0.5) (2p_i(1-p_i)) + (f_i) ((1-p_i)^2) \biggr) \, \prob{f_i | p_i} \, df_i + \right.\\
\left. \qquad + \int_{-\infty}^{0.5} \biggl( (1-f_i) (p_i^2) + (0.5-f_i) (2p_i(1-p_i)) + (f_i) ((1-p_i)^2) \biggr) \, \prob{f_i | p_i} \, df_i  \right] \prob{p_i} \, dp_i 
\label{int3}
\end{multline}
Expanding the polynomials in Eq.~\ref{int3} and once more using Eq.~\ref{defna} to simplify notation, we rewrite the above as
\begin{multline}
\int_{-\infty}^{\infty} \frac{A_{F,i}}{\sqrt{\pi}} \left[ \int_{0.5}^{\infty} \bigl(C_1 f_i + C_2 \bigr)  e^{-A_{F,i}^2 (f_i-p_i)^2} df_i + \right. \\
\left. +  \int_{-\infty}^{0.5} \bigl(C_3 f_i + C_4 \bigr) e^{-A_{F,i}^2 (f_i-p_i)^2} df_i  \right] \prob{p_i}\; dp_i 
\label{int4}
\end{multline}
where $C_1, C_2, C_3,$ and $C_4$ are functions of $p_i$ but independent of $f_i$:
\begin{align}
C_1 &= 1-2p_i^2 \, ,\\
C_2 &= 2p_i^2 -p_i \, ,\\
C_3 &= 1 - 4p_i + 2 p_i^2 \, ,\\
C_4 &= p_i \, .
\end{align}
Performing the interior integration in Eq.~\ref{int4} yields
\begin{multline}
\int_{-\infty}^{\infty} \frac{A_{F,i}}{\sqrt{\pi}}\; \Biggl[ 
%
%
\bigl(C_1 - C_3)\Biggl(\frac{e^{-A_{F,i}^2 (0.5-p_i)^2}}{2 A_{F,i}^2}\Biggr) + (C_3 p_i + C_4) \Biggl(\frac{ \sqrt{\pi}}{A_{F,i}}\Biggr) + \Biggr. \\ 
\Biggl. + \Bigl((C_1 - C_3)p_i + (C_2 - C_4)\Bigr) \Biggl(\frac{ \sqrt{\pi} \, \mathrm{erfc}\bigr(A_{F,i}(0.5-p_i)\bigl)}{2 A_{F,i}}\Biggr)
\Biggr] \; \prob{p_i}\; dp_i \; .
\label{int5}
\end{multline}
Expanding out the various $C$s as well as $A_{F,i}$, we now have for the first term of $\mean{D_i}$
\begin{multline}
\int_{-\infty}^{\infty}  \bigl( p_i(1-p_i) \bigr) \; \Biggl[ 
2 \sqrt{\frac{p_i(1-p_i)}{\pi\; n_F}} \exp{ \Biggl( - \frac{ n_F (0.5-p_i)^2 }{ p_i(1-p_i) } } \Biggr) + \\ + 2 (1-p_i) + 
(2p_i-1) \mathrm{erfc} \Biggl( \sqrt{ \frac{ n_F (0.5-p_i)^2 }{ p_i(1-p_i) } }\Biggr)
\Biggr] \; \prob{p_i}\; dp_i \; ,
\label{int6}
\end{multline}
which has an indirect dependence on $n_F$.  Performing the same integration for the second term in Eq.~\ref{int1} yields analogous indirect $n_G$ dependence.  As a result, when $n_F < n_G$, the first term is greater than the second, yielding $\mean{D_i}>0$; in the limit  $n_F, n_G \rightarrow \infty$, this difference becomes smaller.


The dependence is illustrated in Fig.~\ref{AppFig1}A.  Here, we assume a
uniform distribution of $p_i$ on $(0,0.5)$ and construct $10^5$
$p_i$'s for the underlying population $P$ from which we draw,
independently, a sample $G$ of size $n_G=1000$ and 200 samples $Y$
from which we estimate $\mean{D_i}$ under the null hypothesis.  Sample
$F$ is drawn i.i.d.  from $P$ with sample sizes ranging from $n_F=10$
to $n_F=1000$, permitting us to plot $\mean{D_i}$ as $n_F$ is varied.
The simulation results are shown as circles, overlayed with a plot
of Eq.~\ref{int1} using the result in Eq.~\ref{int6} and  assuming
the uniform distribution of $p_i$.  The values for $\mean{D_i}$ obtained
from the simulation closely matches those derived from Eq.~\ref{int6}.
In Fig.~\ref{AppFig1}B, the corresponding values of $T$ are presented.

\section*{Acknowledgments}
This research was supported by the Intramural Research Program of the
National Cancer Institute, National Institutes of Health, Bethesda, MD.
RB was supported by the Cancer Prevention Fellowship Program, National
Cancer Institute, National Institutes of Health, Bethesda, MD.


\bibliographystyle{unsrt} 
\bibliography{addtobib}


\begin{landscape}
\begin{table}
{\footnotesize{
\begin{tabular}{r|r|r|r}
$Y$ individuals &$F$ population&$G$ population&$T$ distribution \\
\hline
1145 CGEMS cases & \multirow{3}{*}{60 unrelated HapMap CEPH} & \multirow{3}{*}{1145 CGEMS cases} & \multirow{3}{*}{Fig.~\ref{plotdists}} \\
1142 CGEMS controls & & & \\
S.1 -- S.5  & & & \\
\hline
1145 CGEMS cases & \multirow{3}{*}{60 unrelated HapMap CEPH} & \multirow{3}{*}{1142 CGEMS controls} & \multirow{3}{*}{Fig.~\ref{plotdists}} \\
1142 CGEMS controls & & & \\
S.1 -- S.5  & & & \\
\hline
1145 CGEMS cases & \multirow{3}{*}{1142 CGEMS controls} & \multirow{3}{*}{1145 CGEMS cases}& \multirow{3}{*}{Fig.~\ref{plotdists}} \\
1142 CGEMS controls & & & \\
S.1 -- S.5  & & & \\
\hline
100 CGEMS cases not in $G$ & \multirow{4}{*}{1042 CGEMS controls} & \multirow{4}{*}{1045 CGEMS cases}& \multirow{4}{*}{Fig.~\ref{nulldistiswrong}} \\
100 CGEMS controls not in $F$ & & & \\
90 HapMap CEPH & & & \\
90 HapMap YRI & & & \\
\hline
HapMap YRI mothers 16--30 &  \multirow{2}{*}{ HapMap YRI mothers 1--15 and fathers 1--15} &  \multirow{2}{*}{ HapMap YRI children 1--15 and fathers 16--30}& \multirow{2}{*}{Fig.~\ref{hapclass2}}\\
HapMap YRI children 16--30& & & \\
\hline
HapMap CEPH mothers 16--30 &  \multirow{2}{*}{ HapMap CEPH mothers 1--15 and fathers 1--15} &  \multirow{2}{*}{ HapMap CEPH children 1--15 and fathers 16--30}& \multirow{2}{*}{Fig.~\ref{hapclass2}}\\
HapMap CEPH children 16--30& & & \\
\hline
\end{tabular}
}}
\caption{Summary of tests performed.  In the last four rows, the numbers refer to the families in the HapMap YRI and CEPH populations, such that child 1 is the offspring of mother 1 and father 1, et cetera.}
\label{testtab}
\end{table}
\end{landscape}

\clearpage

\begin{table}

\center{
\begin{tabular}{r|r|r|r|r|}
& \multicolumn{2}{c|}{481,382 SNPs} & \multicolumn{2}{c|}{50,000 SNPs} \\
 & $\alpha=0.05$ &  $\alpha=10^{-6}$ & $\alpha=0.05$ &  $\alpha=10^{-6}$ \\
\hline
Sensitivity & 99.8\% & 97.5\% & 96.3\% & 36.3\% \\
Specificity, 200 CGEMS & 31.0\% & 70.5\% & 79.0\% & 99.5\% \\
Specificity, 90 HapMap CEPH &
5.5\% &
27.7\% &
45.5\% &
100.0\% \\
Specificity, 90 HapMap YRI&
0.0\% &0.0\% &4.4\% &97.7\% \\
\hline
\end{tabular}
}
\caption{Empirical sensitivity and specificity for the tests shown in Fig.~\ref{nulldistiswrong} assuming $\mu_0=0$.  Classification results are given for two different  nominal false positive rates $\alpha=0.05$ and $\alpha=10^{-6}$.}
\label{restab}
\end{table}

\begin{table}
\center{
\begin{tabular}{r|r|r|r|r|}
& \multicolumn{2}{c|}{481,382 SNPs} & \multicolumn{2}{c|}{50,000 SNPs} \\
 & $\alpha=0.05$ &  $\alpha=10^{-6}$ & $\alpha=0.05$ &  $\alpha=10^{-6}$ \\
\hline
Sensitivity &
99.90\% &
99.23\% &
97.36\% &
31.09\% \\
Specificity, 200 CGEMS &
40.0\% &
87.0\% &
78.0\% &
99.5\% \\
Specificity, 90 HapMap CEPH &
14.4\% &
55.5\% &
54.4\% &
100.0\% \\
Specificity, 90 HapMap YRI &
0.0\% &
0.0\% &
7.7\% &
100.0\% \\
\hline
\end{tabular}
}
\caption{Empirical sensitivity and specificity for the tests shown in Fig.~\ref{nulldistiswrong} using $\mu_0$ as given by Eq.~\ref{mu05} and assuming that $p_i = (n_F\cdot f_i + n_G\cdot g_i)/(n_F+n_G)$. Classification results are given for two different  nominal false positive rates $\alpha=0.05$ and $\alpha=10^{-6}$.}
\label{restab2}
\end{table}


\newcommand{\freqdistscap} {
Distribution of minor allele frequencies (left) and differences (right)
in CGEMS cases vs HapMap CEPHs (top),  CGEMS controls vs HapMap CEPHs
(center), and  CGEMS cases vs CGEMS controls (bottom).  Note that the
distribution of MAF differences is much narrower when comparing CGEMS
cases to controls (bottom) than when comparing either to HapMap CEPH.
Only SNPs achieving frequencies of 0.05 or more were considered.
}

\newcommand{\AppFigcap} {
Observed $\mean{D_i}$ and $T$ values for simulated data with varying
sample sizes of $n_F$ under the $\mu_0=0$ assumption.  In A, open
circles represent the average $\mean{D_i}$ for each simulation; the solid
line is the theoretical $\mean{D_i}$ based on numerical integration of
Eq.~\ref{int6}. In B, boxplots of the observed $T$s for each simulation
are given assuming $\mu_0=0$; box boundaries correspond to the 0.25 and 0.75 quantiles,
and whiskers indicate the 0.05 and 0.95 quantiles ($T$ values outside
those limits are shown as square points).  Horizontal lines at $T=0$
(green),  $T=1.64$ (corresponding to $\alpha=0.05$, in amber), and
$T=4.75$ (corresponding to $\alpha=10^{-6}$, in red) are shown for
reference; note that for $n_F<600$,  at least 25\% of null samples yield
significant $T$ at the nominal $\alpha=0.05$.  
}

\newcommand{\plotdistscap} {
Distribution of $T$ for real CGEMS samples (left column) and simulated samples
S.1--S.5 (right column) using $F$/$G$ pairs as follows: top,  $F=$ HapMap CEPHs,
$G =$ CGEMS cases; center, $F=$ HapMap CEPHs, $G =$ CGEMS controls;
bottom, $F=$ CGEMS controls, $G =$ CGEMS cases.  Only SNPs achieving
frequencies of 0.05 or more were considered. Note that $\abs{T}>1.64$ is
significant at the nominal $\alpha=0.05$ level and $\abs{T}>4.75$ is
significant at the nominal $\alpha=10^{-6}$ under the putative null distribution.
} 

\newcommand {\nulldistiswrongcap} {
Comparison of $T$ distributions for true positive and negative samples
vs. putative null, starting with 481,382 SNPs in (A,B) and 50,000 SNPs in (C,D).
In all plots, true positive $F$ (1042 CGEMS controls) is shown as a
solid green curve, true positive $G$ (1045 CGEMS cases) is shown as a
solid red curve, and the putative null $N(0,1)$ is given as a thin grey curve.
The dark and light grey regions represent the areas for which the null
hypothesis would be accepted at $\alpha=0.05$ and $\alpha=10^{-6}$,
respectively.  In plots (A,C), CGEMS test samples in neither $F$ nor $G$
(100 CGEMS cases and 100 CGEMS controls) are given by a heavy black curve.
The CGEMS case and CGEMS control distributions within this group are shown as dashed red
and green lines, respectively.  In plots (B,D), $T$ distributions are
given for HapMap CEPHs (cyan) and YRIs (blue).  Vertical lines mark the
0.05 and 0.95 quantiles of the negative CGEMS samples (black),  HapMap
CEPHs (cyan), and HapMap YRIs (blue).
}

\newcommand {\ppvplotcap} {
Positive predictive value (PPV) as a function of prevalence and 
specificity given 99\% sensitivity.  In (A), PPV is shown on the $y$ 
axis and color corresponds to specificity.  The black curve depicts the
87\% sensitivity line---the best sensitivity obtained in the empirical
tests in Tables~\ref{restab},~\ref{restab2}.  In (B), PPV is shown by color,
and the $y$ axis corresponds to specificity.
} 

\newcommand{\hapclasscap}  {
Distributions of $T$ for  out-of-group samples who are related (red line) and unrelated (blue line) to individuals in $G$ for HapMap YRI (A) and HapMap CEPH (B) populations.  (C) and (D) show the same distributions as (A) and (B) respectively, with the addition (green line) of individuals who are in $G$ and unrelated to $F$ (i.e., true positives).  Dashed black lines indicate the $T$ significance thresholds of $\pm 1.64$ at nominal  $\alpha=0.05$.
}

\newcommand{\misclasscap} {
Misclassification rates for samples resembling true positives, as described in
Sects.~\ref{sim2}.  In (A), samples were generated
which had fractional genotype identity to a specific true positive;
the false positive rate is given as a function of the pairwise
similarity.  In (B), samples drawn on a distribution that is a
proportional mixture of $g_i$ and the reference population MAFs; the
false positive rate is given as a function of the proportion of $g_i$.
}

\begin{figure}[htb]
\centering
\includegraphics[width=6in]{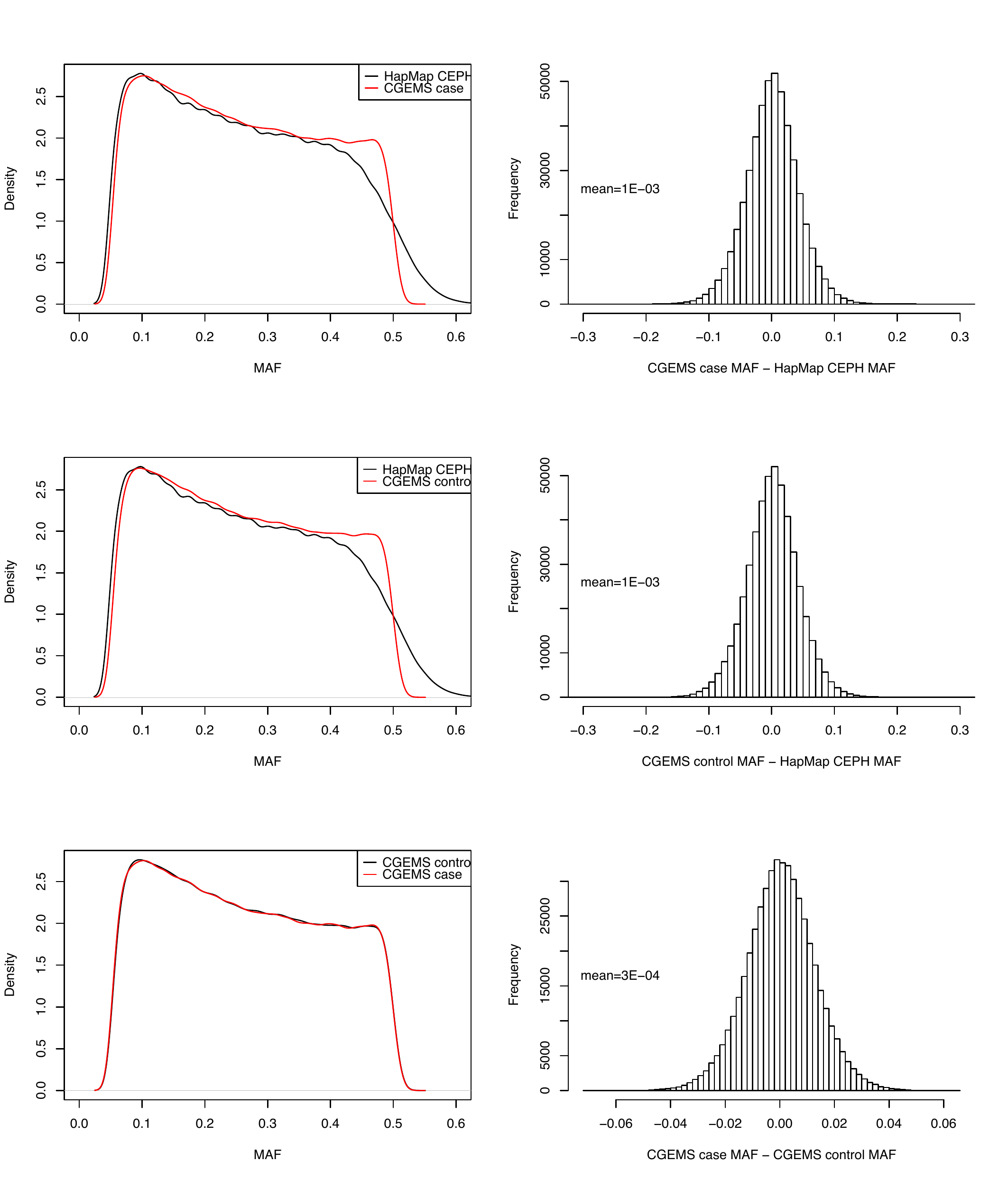}
\caption{\label{freqdists} \freqdistscap}
\end{figure}

\begin{figure}[htb]
\centering
\includegraphics[width=4in]{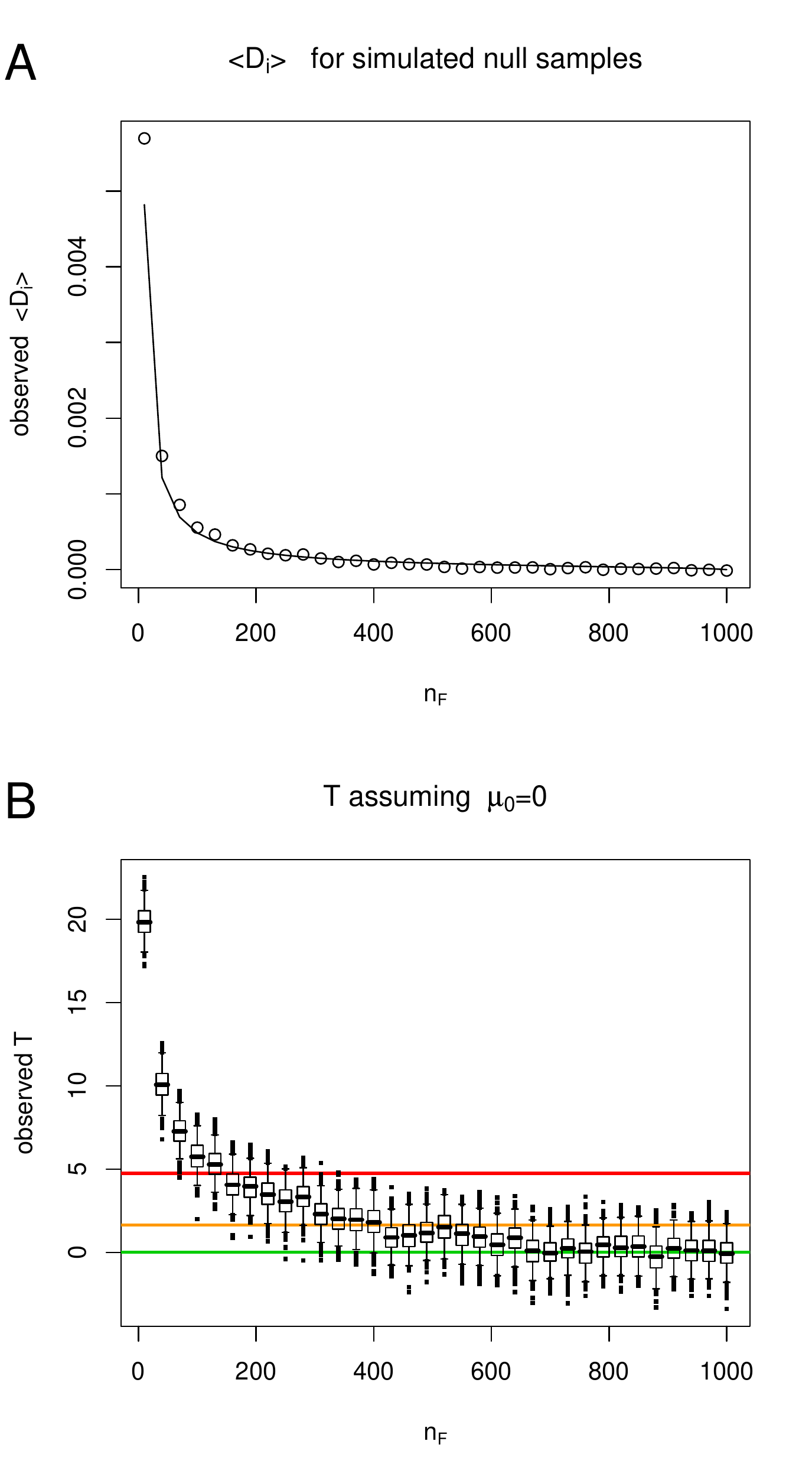}
\caption
{\label{AppFig1} \AppFigcap}
\end{figure}

\begin{figure}[htb]
\centering
\includegraphics[width=6in]{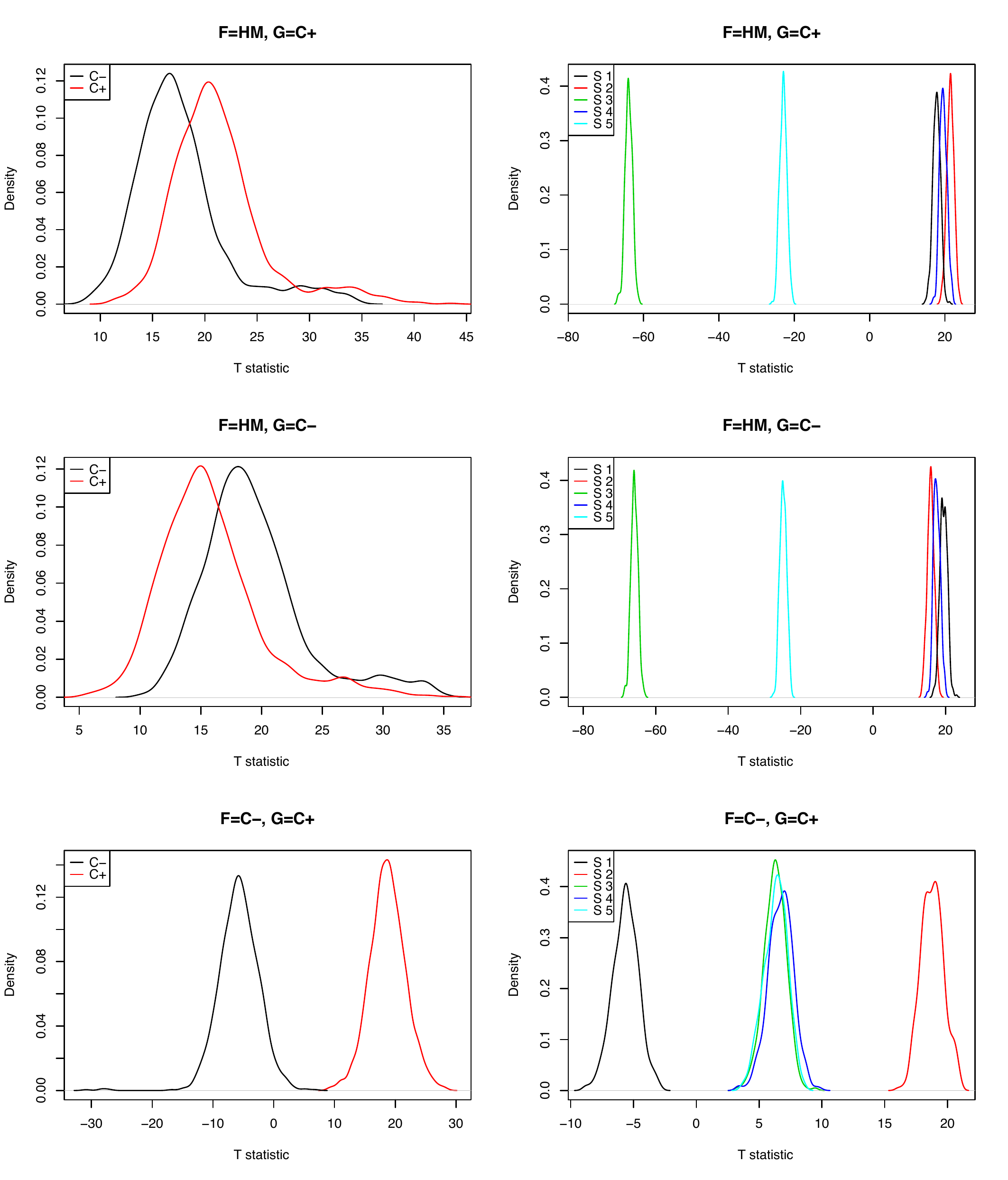}
\caption{\label{plotdists} \plotdistscap}
\end{figure}

\begin{figure}[htb]
\centering
\includegraphics[width=6in]{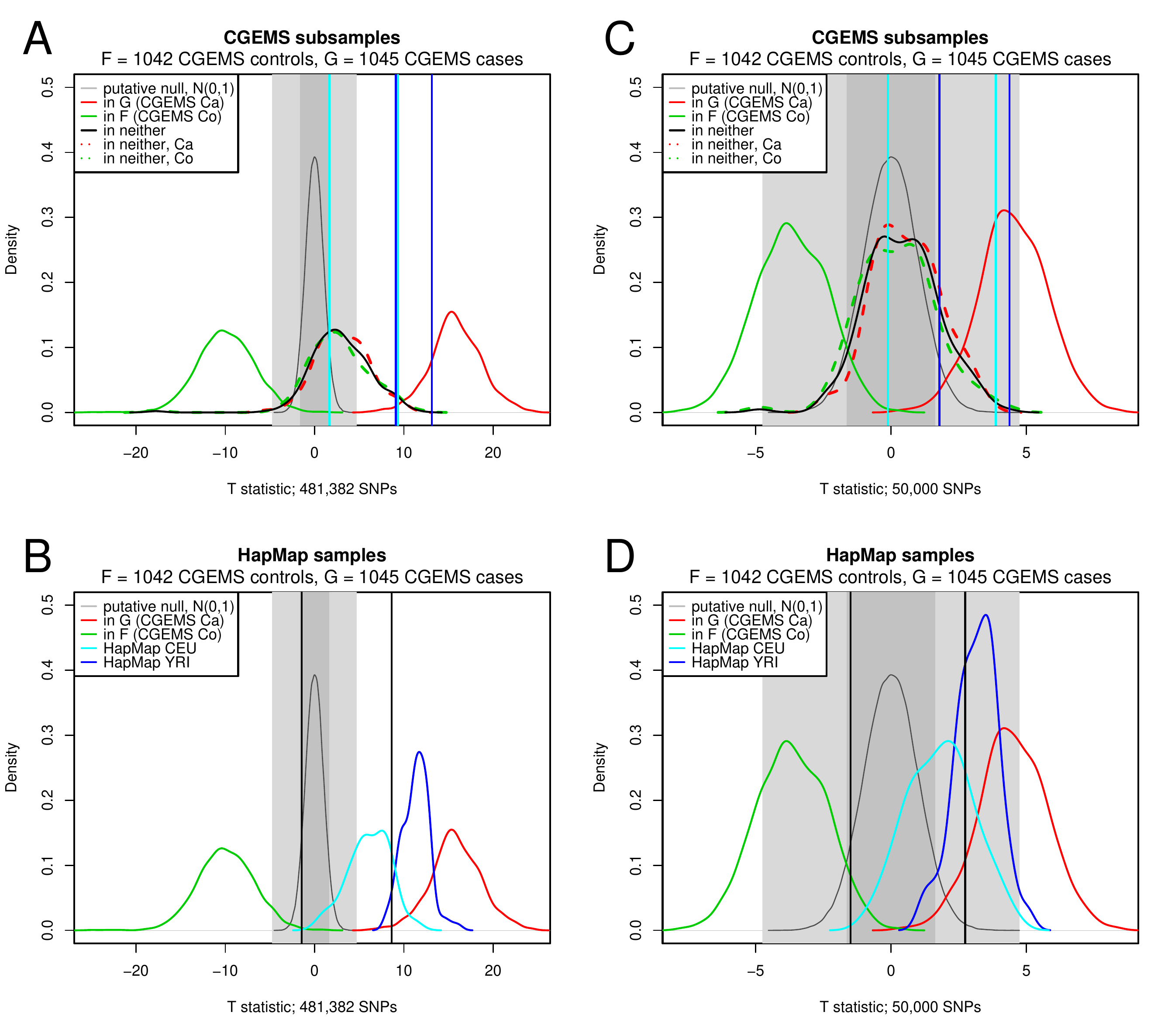}
\caption{\label{nulldistiswrong} \nulldistiswrongcap}
\end{figure}

\begin{figure}[htb]
\centering
\includegraphics[width=3in]{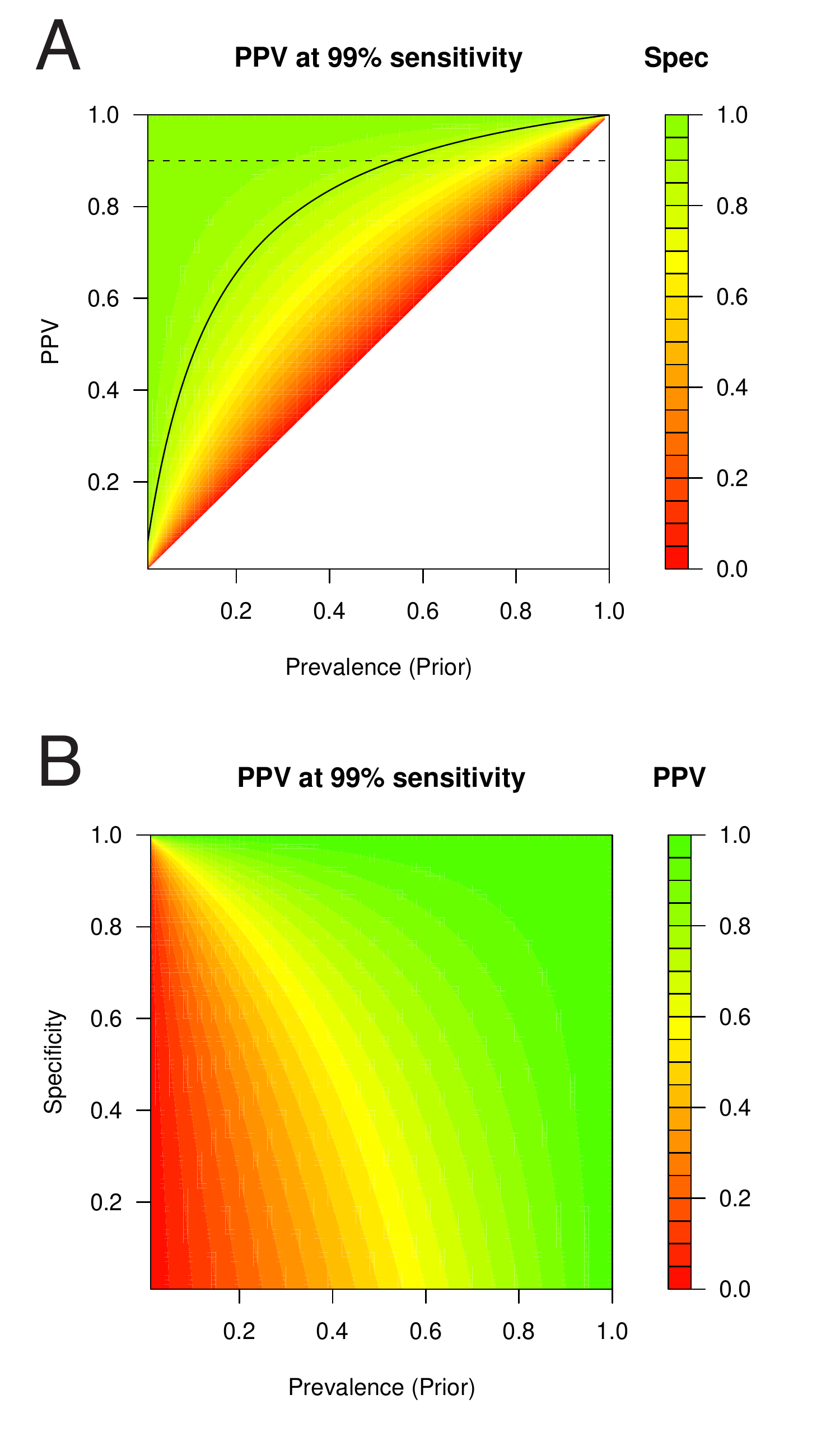}
\caption
{\label{ppvplot} \ppvplotcap}
\end{figure}

\begin{figure}[htb]
\centering
\includegraphics[width=6in]{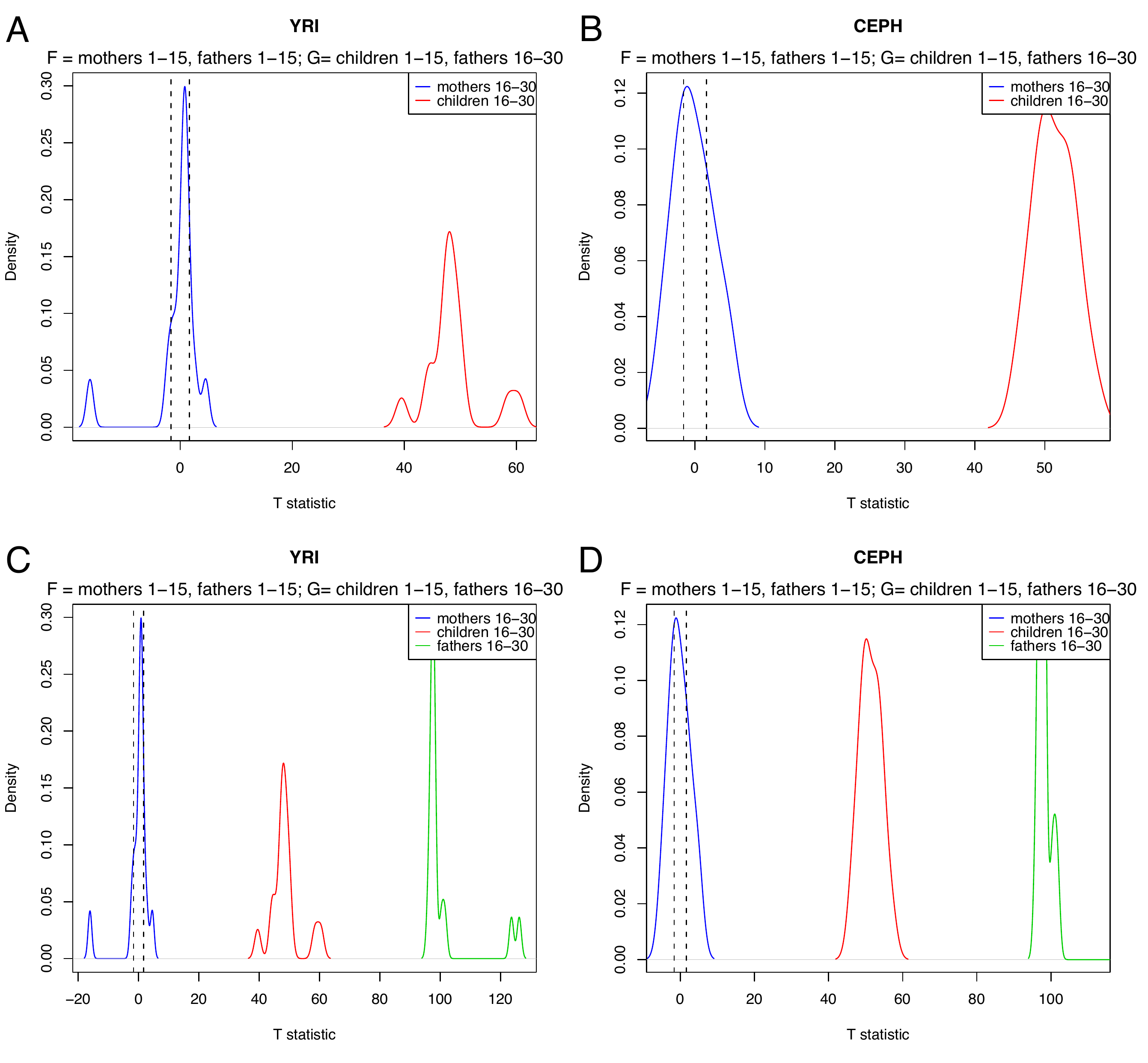}
\caption{\label{hapclass2} \hapclasscap}
\end{figure}

\begin{figure}[htb]
\centering
\includegraphics[width=4in]{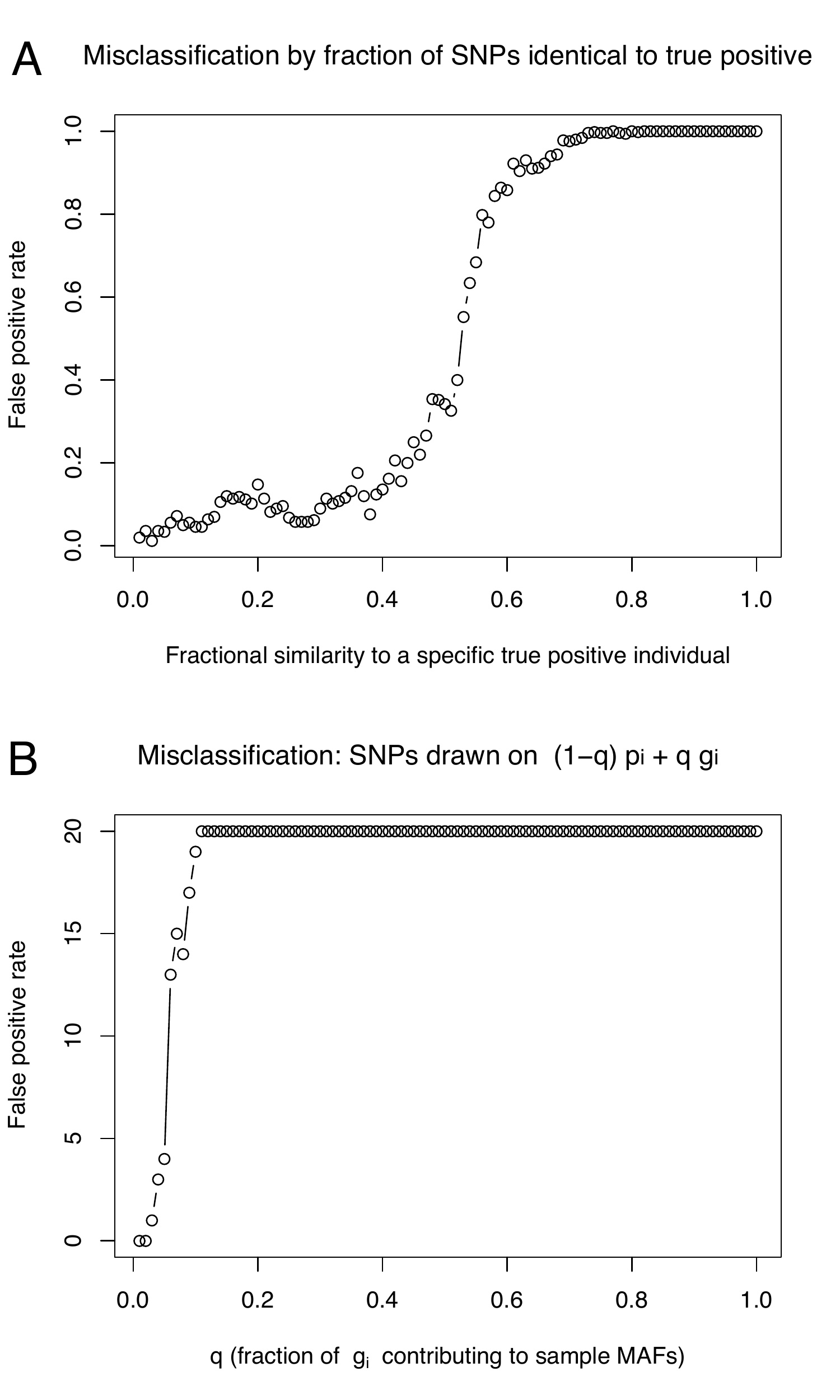}
\caption{\label{misclass} \misclasscap}
\end{figure}

\end{document}